\documentclass[aps,prd,groupedaddress,twocolumn]{revtex4}
\usepackage{graphicx,subfigure}
\usepackage{dcolumn}
\usepackage{bm,amsfonts,amssymb,amsmath}
\usepackage{color}
\usepackage{longtable}
\usepackage{booktabs}
\usepackage{multirow}
\usepackage[breaklinks=true,colorlinks=true,linkcolor=blue,
urlcolor=blue,citecolor=blue]{hyperref}

\def\bea{\begin{eqnarray}}
\def\eea{\end{eqnarray}}
\def\l{\left}
\def\rt{\right}

\def\nn{\nonumber}
\def\Eq#1{Eq.~(\ref{#1})}
\def\Eqs#1{Eqs.~(\ref{#1})}
\def\Fig#1{Fig.~\ref{#1}}
\def\Figs#1{Figs.~\ref{#1}}
\def\abs#1{\left|#1\right|}
\def\xk#1{\left(#1\right)}
\def\zk#1{\left[#1\right]}

\def\Im{{\rm Im}}

\def\trace#1{{\rm Tr}\left[#1\right]}
\def\Det#1{{\rm Det}\left[#1\right]}

\def\pa{\partial}
\newcommand{\s}{{\sigma}}
\newcommand{\Si}{\Sigma}

\newcommand{\de}{\delta}
\newcommand{\De}{\Delta}
\newcommand{\ep}{\epsilon}
\newcommand{\ga}{\gamma}
\newcommand{\Ga}{\Gamma}

\newcommand{\la}{\lambda}
\newcommand{\La}{\Lambda}
\newcommand{\om}{\omega}

\renewcommand{\th}{\theta}

\renewcommand{\b}[1]{{\bf #1}}

\begin{document}

\title{Interplay between tilt, disorder, and Coulomb interaction in type-I Dirac fermions}

\author{Peng-Lu Zhao}
\altaffiliation{zpljdwlx@mail.ustc.edu.cn}\affiliation{Department of Modern Physics, University of Science and Technology of China, Hefei, Anhui 230026, P. R. China}
\author{An-Min Wang}
\affiliation{Department of Modern Physics, University of Science and
Technology of China, Hefei, Anhui 230026, P. R. China}

\begin{abstract}
We investigate the mutual influence of tilt, disorder, and Coulomb interaction
in a type-I Dirac semimetal (DSM) with $x$-direction tilt by performing a
renormalization group analysis. The interplay between disorder and ordinary
tilt generates an effective tilt along the $x$-direction, which is the physically
observable one. There exist two types of disorder which increase the effective
tilt and drive a phase transition from the DSM phase to the diffusive metal phase.
The diffusive phase transition stops the increase of the effective tilt and
the surface of the original Dirac cone in the diffusive metal phase is just
slightly tilted. Surprisingly, the Dirac point is replaced by a bulk nodal arc
in the diffusive metal phase. The Coulomb interaction suppresses the diffusive
phase transition and therefore is harmful to the formation of bulk nodal arc.
In contrast, there also exists other two types of disorder which reduce the
effective tilt and induce no phase transition. For these two types of disorder,
the Coulomb interaction enhances their low-energy relevances. Coexistence of
Coulomb interaction with any of them leads to a stable infrared fixed point where
the coupling strengths for two kinds of interaction are identical and the effective
tilt vanishes. The original tilted Dirac semimetal now reacts like an untilted and
interaction-free Dirac semimetal. Our results show that interplay between tilt,
disorder, and Coulomb interaction results in rich low-energy properties for
the tilted Dirac fermions.
\end{abstract}


\maketitle

\section{Introduction}

Recently, a new class of Dirac/Weyl fermions with a tilted
conic spectrum \cite{Katayama06,Kobayashi07,Goerbig08,Kobayashi08},
called tilted Dirac/Weyl fermions, has received considerable theoretical and
experimental interest in the condensed matter community.
It has been found that the tilt plays a key role in this new class of
Dirac fermions. When the effective tilt is sufficiently large,
the Fermi surface crossing the Dirac nodes becomes lines in two
dimensions \cite{Katayama06,Kobayashi07,Goerbig08,Kobayashi08}
and a surface in three dimensions \cite{Soluyanov15}.
Such a system is usually identified as a type-II Dirac/Weyl semimetal (D/WSM)
\cite{Soluyanov15,YSun15,Koepernik16,Autes16}.
Several novel phenomena have been predicted to arise due to this unusual
Fermi surface, including an unconventional magnetic-optical response
\cite{Proskurin15, ZMYu16,Tchoumakov16,Udagawa16}
and magnetic breakdown \cite{OBrien16}, anomalous Hall effect\cite{Zyuzin16, Steiner17},
and anomalous Nernst and thermal Hall effects \cite{Ferreiros17, Saha18}.
Moreover, several proposals have been presented for the realization of
such tilted Dirac/Weyl fermions in different regimes
\cite{Katayama06,Kobayashi07,Goerbig08,Kobayashi08,Soluyanov15, Yxu15, Yxu16},
and some experiments \cite{LNHuang16, CLWang16, MZYan17, Noh17, KNZhang17}
performed via angle-resolved photoemission spectroscopy have reported evidence
for the existence of type-II Dirac/Weyl fermions.

After the discovery of type-II Dirac fermions, the Lifshitz transition, which is identified by a change in the topology of the Fermi surface, between different types of Dirac fermions has attracted
considerable attention. It was first suggested that Coulomb interaction decreases the tilt and may lead to a transition from type-II Dirac fermions to type-I ones \cite{Isobe16}. Subsequent researches \cite{ZMHuang17,YWLee18} show that Coulomb interaction is completely screened in type-II DSMs due to the nonvanishing density of states (DOS) at the Fermi level. The screened Coulomb interaction cannot reduce the tilt. As a result, it stabilizes the type-II Dirac fermions. However, for a type-I DSM which still has pointlike Fermi surface, the vanishing DOS implies that the Coulomb interaction is not screened sufficiently \cite{Kotov12}. As a consequence, the Coulomb interaction reduces the tilt for type-I Dirac fermions \cite{Isobe12,YWLee18}.
In addition, it has been found that the electromagnetic field in a  three-dimensional (3D) WSM generates a
stable infrared fixed point with a finite value for tilt \cite{Pozo2018}.

The disorder is another factor which has a substantial influence to the tilt. In specific, charge and magnetic disorder can increase the tilt by reducing the topological mass \cite{YJWu17, Park17} of tilted Dirac/Weyl SMs. With tilt increasing, the transition from type-I to type-II Dirac/Weyl fermions appears. The nonmagnetic disorder in an isolated Weyl cone also enhances the tilt \cite{Trescher17, Sikkenk17}. At the same time, it drives a phase transition from the SM phase to compressible diffusive metal (CDM) phase \cite{Trescher17, Sikkenk17}, which is characterized by the generation of finite zero-energy
scatting rate $\ga_0$ and finite zero-energy density of states (DOS) $\rho\xk{0}$ \cite{Trescher17, Sikkenk17,Fradkin1, Fradkin2,Shindou09,Ominato14,Goswami11,Kobayashi14,Sbierski14,Roy14,
Syzranov15L,Syzranov15B,Pixley15}. This disorder-induced SM-to-CDM phase transition has been widely studied in un-tilted D/WSMs \cite{Fradkin1, Fradkin2,Shindou09,Ominato14,Goswami11,Kobayashi14,Sbierski14,Roy14,
Syzranov15L,Syzranov15B,Pixley15}. In addition, the disorder also induces the well known metal-insulator transition for tilted Weyl fermions \cite{YJWu17,Abrahams79,MacKinnon81}. A more interesting disorder-induced phenomenon in tilted Dirac/Weyl SMs is that some special kinds of disorder can disrupt the Dirac point and produce a bulk nodal arc \cite{Papaj18,PLZhao2018,Carlstroem2018,Carlstroem2019}. So far, this novel phenomenon is only reported to arise in a tilted DSM which breaks any external symmetry \cite{Papaj18,PLZhao2018}.

Based on previous findings that Coulomb interaction decreases the tilt \cite{Isobe12,Detassis17,ZMHuang17,YWLee18} while disorder increases it \cite{Park17,Sikkenk17}, the fate of tilt is highly dependent on the interplay of these two kinds of interaction. Moreover, when Coulomb interaction coexists with disorder, a variety of intriguing properties appear due to their interplay\cite{Finkelstein84, Castellani84,Abrahams01,Kravchenko04, Punnoose05,Spivak10, Ye98, Ye99,Stauber05, Herbut08, Vafek08, Foster08,Goswami11, WangLiu14, PLZhao16, Roy16, Gonzalez17, Nandkishore17, YXWang17}. This issue has been known to play a vital role in
determining the properties of two-dimensional (2D) metallic systems
\cite{Finkelstein84, Castellani84, Abrahams01,Kravchenko04,Punnoose05, Spivak10} and has been extensively studied for over three decades. In this paper, we perform a systematic renormalization group (RG)
investigation for the influence of Coulomb interaction and disorder to the stability
of tilted type-I Dirac fermions in two dimensions.
For simplicity, we consider a tilt along the $x$-direction while the $y$-direction
is not tilted. We mainly focus on the three types of disorder which are extensively
studied in untilted Dirac semimetals \cite{Ludwig94,Ye99,Stauber05,Ostrovsky2006,Herbut08,Goswami11}.
Namely, the random scalar potential (RSP), the random
vector potentials (RVP), and the random mass (RM) \cite{Ludwig94,Ye99,Stauber05,Ostrovsky2006,Herbut08,Goswami11}.
Moreover, in consideration of the anisotropy of the Dirac cone,
we distinguish the two components of RVP by their directions and
denote them as $x$-RVP and $y$-RVP, respectively. Our main findings are as follows.

(1) In the formalism of RG, the interplay between ordinary tilt and disorder dynamically generates an anomalous tilt, and the interplay of disorder and Coulomb interaction gives rise to a new type of Coulomb vertex automatically. After including the effects of these dynamically generated terms by the way provided in Ref. \cite{Sikkenk17}, we found a definition of effective tilt which represents the physically observable tilt and can be perceived by DOS and specific heat. By determining the fate of the effective tilt under the influence of Coulomb interaction and disorder, we find that the Lifshitz transition between two types of Dirac fermions never appears although the specific way to achieve this result is dependent on the disorder type.

(2) Without Coulomb interaction, the RSP and $x$-RVP both enhance the effective tilt, which means they are able to cause the phase transition between type-I and type-II DSMs. At the same time, they induce the diffusive phase transition and this transition precedes the Lifshitz transition. As a result, the tilt is much less than one and the surface of original Dirac cone is just slightly tilted in the CDM phase. In addition, a bulk nodal arc arises and replaces the Dirac point, which phenomenon has been reported in Refs. \cite{Papaj18,PLZhao2018} previously. For $y$-RVP and RM, they directly reduce the effective tilt and can not induce the diffusive phase transition. Therefore, the type-I Dirac fermions are stable in the disordered DSM phase. In addition, we found that the ordinary tilt along the $x$-direction breaks down the local gauge symmetry depending solely on the $x$-coordinate and promotes $x$-RVP to become marginally relevant. However, there is no tilt along the $y$-direction, which keeps the local gauge symmetry depending solely on the $y$-coordinate intact and guarantees $y$-RVP to be marginal at any order of loop expansion.
Moreover, we also calculated the low-energy behavior of DOS and specific heat. The results show that RSP and $x$-RVP both generate a nonzero zero-energy DOS and a linear temperature dependence for specific heat. The $y$-RVP gives rise to a power-law enhancement to DOS and specific heat. The DOS and specific heat are also logarithmically enhanced by RM.

(3) When Coulomb interaction is present, it always reduces the effective tilt.
The interplay of Coulomb interaction and RSP generates a critical RSP strength
to produce the CDM transition, as a result, the DSM and CDM phases are separated
by a line in the plane formed by these two interaction couplings and tilt is
much less than one for both phases. However, there is no bulk nodal arc in the
DSM phase, which indicates the Coulomb interaction will destroy the RSP induced
bulk nodal arc if the given strength of RSP is weaker than the critical one.
In addition, with the ordinary tilt increasing, the value of critical RSP strength
also increases, which means ordinary tilt is helpful to stabilize the DSM phase.
The interplay of Coulomb interaction with any components of RVP or RM generates
a randomly stable state which is characterized by the existence of a common
constant value for the coupling strengths of all interactions in the low-energy limit
and the constant value is not fixed, its value is highly random and dependent on
the initial conditions of the system. For this state, the effective tilt and
velocity anisotropy vanish in the low energy limit, the type-I Dirac fermions
keep stable, the DOS and specific heat display the same asymptotic behavior as
the free system in low energy. The bulk nodal arc is absent for this state and,
therefore, we conclude that Coulomb interaction turns the $x$-RVP produced bulk
nodal arc into a Dirac point.

The remainder of the paper is organized as follows. In Sec.~\ref{Sec:Model}, we present
the model of the 2D tilted Dirac fermion system and derive the RG equations. We analyze the impact of
every single type of random potential in Sec.~\ref{Sec:Nl}. The interplay of Coulomb interaction with random potentials is presented in Sec.~\ref{Sec:IwCi}.
We summarize the results and address a related issue in Sec.~\ref{Sec:Summ}.

\section{The setup \label{Sec:Model}}

In this section, we present the low energy effective model and derive the full set of RG equations as a bias to proceed.

\subsection{Effective action}

We start with the minimal model of noninteracting tilted
Dirac fermions near a single Dirac cone with chirality $\chi=+1$ described by the Bloch Hamiltonian \cite{Goerbig08}
\bea
H_0 (\b k) =\psi^{\dag}(\b k)\zk{\xk{t\s_0+\s_1}v_xk_x+v_y\s_2k_y }\psi(\b k),\label{EqfrHam}
\eea
where $\sigma_0$ is the $2\times 2$ identity matrix and the $\sigma_i$ ($i=1,2,3$) correspond to the Pauli matrices. $v_x$ and $v_y$ denote
the velocity along the $x$ and $y$ directions, respectively. Without
loss of generality, we choose $v_x,v_y > 0$. $t$ represents a dimensionless
tilting parameter along the $x$ axis. For type-I Dirac fermions, the tilt is limited to $\abs{t}<1$, and for type-II, $\abs{t}>1$. $\abs{t}=1$ correspond to the Lifshitz transition points, which separate the two types of Dirac fermions. In this work, we focus on the stability of type-I Dirac fermions under the influence of Coulomb interaction and disorder. Moreover, when tilt approaches the Lifshitz transition points, the linearized model \Eq{EqfrHam} is not well applied because higher-order corrections become relevant \cite{Ferreiros17,Zyuzin18}. Therefore, the tilt is restricted to $\abs{t}\ll 1$ in this work. The energy dispersion of the Hamiltonian in \Eq{EqfrHam} reads as follows:
\bea
{\cal E}_{\pm}\xk{\b k}=tv_{x}k_{x}\pm\sqrt{v_{x}^{2}k_{x}^{2}+v_{y}^{2}k_{y}^{2}}.\label{Eqdisp}
\eea
Due to $\abs{t}\ll 1$, the fermion surface still keeps pointlike.

We notice that another model of tilted Dirac fermion was put forward in Ref. \cite{Papaj18}, wherein the two components of Dirac fermions are composed of two distinct degrees of freedom and thus unrelated by any symmetry. However, for our model, the two components of Dirac fermions refer to the real spin degree of freedom and they are related by the ``particle-hole" symmetry \cite{YWLee18, Li2016} as
\bea
\mathcal{P}_h^{-1}H_0 ({ \b k})\mathcal{P}_h =-H_0^{\ast} (-{ \b k}),\label{EqphtranH0}
\eea
where the ``particle-hole'' transformation is implemented by a unitary operator $\mathcal{P}_h$, which is defined by $\mathcal{P}_h^{-1}\psi(\b{k})\mathcal{P}_h=\s_1\psi^{\ast}(-\b{k})$.

The long-range Coulomb interaction between fermions is given by
\bea
H_C=\frac{1}{2}\int d^2\b xd^2\b y\rho(\b{x})\frac{e^2}{4\pi\ep\abs{\b x-\b y}}\rho(\b{y}), \label{EqCoulHam}
\eea
where $\rho(\b{x})=\psi^{\dagger}(\b{x})\psi(\b{x})$ is the
normal-ordered electron density operator, $\ep$ is the dielectric constant, and $-e$ is the
charge carried by an electron. After performing a standard Hubbard-Stratonovich transformation, the action for Coulomb interaction in the imaginary-time formulation can be written as
\bea
S_C =\int d\tau d^2 x \xk{ig\psi_a^\dagger \phi \psi_a}
+ \frac{1}{2}\int d\tau d^3 x \!\!\sum_{j=x,y,z} \l(\partial_{j} \phi\rt)^{2},\quad \label{EqCoulAct}
\eea
where $g=e/\sqrt{\epsilon}$ and $\phi$ is a real bosonic field which is introduced through the Hubbard-Stratonovich transformation.

We add the fermion-disorder coupling
to the system by the standard form as \cite{Ludwig94, Nersesyan95,
Altland02, Stauber05},
\begin{eqnarray}
S_{\mathrm{dis}} = \int d^2x d\tau \psi^\dag\left(\sum_{i=0}^3
A_{i}(\mathbf{x})\s_i\right)\psi,\label{EqdisSnor}
\end{eqnarray}
where the function $A_{i}(\mathbf{x})$ stands for the randomly
distributed potential. We assume $A_{i}(\mathbf{x})$ to be a
quenched, Gaussian white-noise potential, which is characterized by the
following identities:
\begin{eqnarray}
\langle A_{i}(\mathbf{x})\rangle = 0,\qquad \langle
A_{i}(\mathbf{x})A_{j}(\mathbf{x}')\rangle =
\kappa_{i}\de_{ij} \delta^2(\mathbf{x}-\mathbf{x}').\label{Eqdefdis}
\end{eqnarray}
Here, the dimensionless variances $\kappa_i$ are introduced to
characterize the strengths of random potentials. The disorder type
is determined by the Pauli matrix. In particular, $
\s_0$ represents RSP, $\s_{z}$ stands for
RM, and $\s_{x},\s_{y}$ denote $x$-RVP and $y$-RVP, respectively. These types of disorder
are most frequently studied and they can be induced by various mechanisms in realistic materials~\cite{Meyer07,CastroNeto09,Peres10,Mucciolo10,Champel10,Kusminskiy11}. Any of these types of random potential might
exist individually. To be general, we assume
they coexist in our next calculation of the RG equations. The random potential $A_i(\mathbf{x})$ needs to be properly averaged. To do this, we assume that the spatial distribution of $A_{i}\xk{\b x}$ is described by $P\zk{A_{i}} =\exp\zk{-\int d^2 \b x A_{i}^2\xk{\b x}/(2\kappa_i)}$.
By employing the most widely used replica method~\cite{Lee85,
Altlandb,Colemanb} to perform the disorder average, we obtain an effective replicated
action in the Euclidean space:
\bea
S_{\mathrm{dis}}=-\frac{1}{2}\int d^2x d\tau
d\tau' \sum_{i=0}^3\kappa_i\big(\psi^{\dagger}_m\s_i
\psi_m\big)_x \big(\psi^{\dagger}_n\s_i
\psi_n\big)_{x'}. \label{Eqactdis}
\eea
Here, $x
\equiv(\mathbf{x},\tau)$ and $x'\equiv(\mathbf{x},\tau')$, $m$ and $n$ are the replica indices which are summed up automatically.

The combination of tilt and disorder [\Eqs{EqfrHam} and (\ref{Eqactdis})]
can not provide self-closed one-loop RG corrections. The
self-energy generates a marginal term proportional to $i\om \s_1$ [see \Eq{Eqdisseen} below], which is not present in the original action.
To keep the RG calculation closed, we add a term of $i\om \la\s_1$ to the
original action by hand, which was suggested in Ref. \cite{Sikkenk17}.
However, this term breaks down the U(1) gauge invariance of the original
action for Coulomb interaction. To repair this, we need to add another
term as $ig \xk{\psi^\dagger_{\alpha}\la\s_1 \psi_{\alpha} }$ in Yukawa coupling.
Even if we do not add this term, the interplay of disorder and Coulomb interaction
will dynamically generate it. With this term, we change the Coulomb vertex to
be $ig \psi^\dagger_{\alpha}\xk{\s_0+\la\s_1 }\psi_{\alpha}$. We stress that
terms of $i\om \la\s_1$ and $ig\xk{\psi^\dagger_{\alpha} \la\s_1 \psi_{\alpha} }$
do not exist in a free Hamiltonian. They all stem from the interplay of disorder
and tilt and they appear only when disorder effect is considered.
As explained in Ref. \cite{Sikkenk17}, the term of $i\om \la\s_1$ is connected to the quasiparticle weight, which has important physical influence. After including
above two terms, the total action in momentum space with Matsubara frequency reads as
\begin{widetext}
\bea
 S&=&
\int \frac{d^2\mathbf{k}d\omega}{(2\pi)^3}\psi^{\dag}_m\zk{-i\om \xk{\s_0+\la \s_1} + v_x t k_x\s_0 +v_x k_x \s_1+v_y k_y \s_2}  \psi_m
+\frac{1}{2}\int \frac{d^2\mathbf{k}d\omega}{(2\pi)^3}\phi\xk{k} D_{0}^{-1}\xk{\om,\b k}  \phi\xk{-k}
\nn\\&&-
\int\frac{d\omega_1
d\omega_2d^{2}\mathbf{k}_1d^{2}\mathbf{k}_2d^{2}\mathbf{k}_3}{(2\pi)^{8}}
\Bigg[\sum_{i=0}^3\frac{\kappa_i}{2}
\psi^\dagger_{m}(i\omega_1,\mathbf{k}_1)\s_i
\psi_{m}(i\omega_1,\mathbf{k}_2)
\psi^\dagger_{n}(i\omega_2,\mathbf{k}_3)\s_i\psi_{n}(i\omega_2,\mathbf{k}_1
-\mathbf{k}_2+\mathbf{k}_3)\Bigg]
\nn \\ &&+\int \frac{d^2\mathbf{k}d\omega}{(2\pi)^3} \frac{d^2\mathbf{k'}d\omega'}{(2\pi)^3}ig
\psi^\dagger_m\xk{\om,\b k}\xk{\s_0+\la \s_1} \psi_m\xk{\om',\b k'}\phi\xk{\om-\om',\b k-\b k'},\label{Eqaction}
\eea
\end{widetext}
where
\bea
D_{0}\xk{\om,\b{k}} = \int \frac{dk_z}{2\pi} \frac{1}{ k_x^2 + k_y^2 + k_z^2}
= \frac{1}{2\sqrt{k_{x}^{2}+  k_{y}^{2}}}.
\eea
This action is the basis of our RG analysis. Under the amendment, the quasiparticle energy dispersion now becomes
\bea
{\cal E}' ({\bf k})&=&\frac{1}{1-\la^2}\bigg[ \xk{t-\la }v_{x}k_{x}
\nn\\&&\pm\sqrt{\xk{1-t\la}^2v_{x}^{2}k_{x}^{2}+\xk{1-\la^2}v_{y}^{2}k_{y}^{2}} \bigg]
\nn\\&=&
t'v'_{x}k_{x}\pm\sqrt{v_{x}^{'2}k_{x}^{2}+v_{y}^{'2}k_{y}^{2}} \label{Eqeigenla},
\eea
where we have defined the effective tilt and velocities to be
\bea
t'=\frac{t-\la }{1-t\la},\quad v_x'=\frac{1-t\la }{1-\la^2}v_x,\quad v_y'=\frac{v_y }{\sqrt{1-\la^2}}\label{Eqeffquan}.
\eea
According to Ref. \cite{Sikkenk17}, these three quantities are identified to be the physically observable tilt and velocities, and they will reduce to the ordinary ones if $\la=0$.

\subsection{RG equations}

We have carried out a detailed RG analysis starting from the action of \Eq{Eqaction}, by considering the leading-order weak-disorder expansion. After integrating out
the fast modes defined within the momentum shell $e^{-\ell}\Lambda <
|\mathbf{p}| < \Lambda$ and then performing RG transformations
\cite{Shankar94}, we obtain the following RG equations:
\bea
\frac{d \la}{d\ell}\!\!&=&\!\!-
\frac{1-\la t}{\xk{1-t^2} }\big[\xk{\De_0+\De_1}\xk{t+\la }
-\xk{\De_2+\De_3} \xk{t-\la } \big],\nn\\ \label{EqRGla}
\\
\frac{d t}{d\ell}\!\!&=&\!\!-\alpha f_x \xk{ t-\la},\label{EqRGt}
\\
\frac{d v_x}{d\ell}\!\!&=&\!\! \zk{\alpha f_x -\frac{ \xk{1-\la t}\xk{\sum_{i=0}^3\De_i}  }{1-t^2}} v_x,\label{EqRGvx}
\\
\frac{d\eta}{d\ell}\!\!&=&\!\! \alpha\xk{ f_y-f_x }\eta,\label{EqRGeta}
\\
\frac{d \alpha}{d\ell}\!\!&=&\!\!-\alpha\zk{\alpha f_x -\frac{ \xk{1-\la t}\xk{\sum_{i=0}^3\De_i}  }{1-t^2}},\label{EqRGalpha}
\\
\frac{d \De_0 }{d\ell}\!\!&=&\!\!\frac{2\De_0\xk{ \sum_{i=0}^3\De_i } }{1-t^2} +2\xk{\Delta_1+\frac{\De_2}{1-t^2}} \Delta_3
\nn\\&&
-\De_0\alpha\zk{f_x\xk{\frac{2\la^2}{1-\la t} + \frac{1-\la t}{1-t^2}  }+ f_y},\label{EqRGDe0}
\\
\frac{d \De_1 }{d\ell}\!\!&=&\!\!
\frac{2t^2\xk{\De_0+\De_1-\De_2-\De_3}\De_1 }{1-t^2}+2 \Delta_3\xk{\frac{t^2\De_2}{1-t^2}+\De_0}
\nn\\&&
+\De_1\alpha \zk{f_x\xk{\frac{2}{1-\la t}-\frac{1-\la t}{1-t^2}  }-f_y },\label{EqRGDe1}
\\
\frac{d \De_2}{d\ell}\!\!&=&\!\!\De_2
 \alpha \zk{f_y -\frac{ f_x\xk{1-\la t}  }{1-t^2}} +\frac{2 \Delta_3\xk{t^2\De_1+\De_0} }{ 1-t^2}, \label{EqRGDe2}
\\
\frac{d \De_3 }{d\ell}\!\!&=&\!\!
-2\De_3\xk{\De_3+\De_0-\De_1-\De_2}
\nn\\&&
+ \frac{2\zk{ \De_0\xk{\De_1+\De_2}+t^2\De_1\xk{\De_2-\De_0} }}{1-t^2}
\nn\\&&
+\De_3\alpha \zk{f_y+f_x\xk{\frac{2\xk{1-\la^2} }{1-\la t}-\frac{ 1-\la t  }{1-t^2}}  }.\label{EqRGDe3}
\eea
where
\bea
f_x&=&\frac{1}{\pi}\int_0^{2\pi}\frac{B\cos^2\th d\th}{ \sqrt{B^2\cos^2\th+\eta^2\sin^2\th}    },
 \nn\\
f_y&=&\frac{1}{\pi}\int_0^{2\pi}\frac{\sqrt{1-\la^2}\sin^2\th d\th}{ \sqrt{B^2\cos^2\th+\eta^2\sin^2\th}    }\label{Eqfxy},
\nn\\
B&=&\frac{1-\la t}{\sqrt{1-\la^2}},\quad \eta=\frac{v_y}{v_x} \label{EqAeta}.
\eea
The detailed derivations are provided in Appendix \ref{Sec:ApenRG}. Here, we have made the
re-definition $\frac{\kappa_i }{2\pi v_x v_y \sqrt{1-t^2}  } \equiv \De_i$, which represent the effective disorder strength. The effective Coulomb interaction is defined by $\alpha \equiv g^2/\xk{16\pi v_x}$. We notice that, by setting $t=\la=\alpha=0$, \Eqs{EqRGvx} and (\ref{EqRGDe0})--(\ref{EqRGDe3}) are in accordance with the results
obtained previously in Refs. \cite{Ostrovsky2006,Aleiner2006,Foster08,Foster2012}, wherein disorder effects for untilted Dirac fermions are well studied.

\section{Noninteracting limit \label{Sec:Nl}}

In this section, we consider the noninteracting limit
by taking $\alpha=0$ and analyze the disorder-induced properties of tilted Dirac fermions. First, we locate the fixed points for the related parameters in the low-energy limit, which can help us to adjudge the relevance (or irrelevance) of each type of disorder. Second, for a relevant disorder which always flows to a strong coupling regime and breaks down the RG, we will show that a bulk nodal arc is produced by using the self-consistent Born approximation (SCBA).

\subsection{Fixed points \label{Sec:Fp}}

\begin{figure}[htbp]
\center
\includegraphics[width=3.3in]{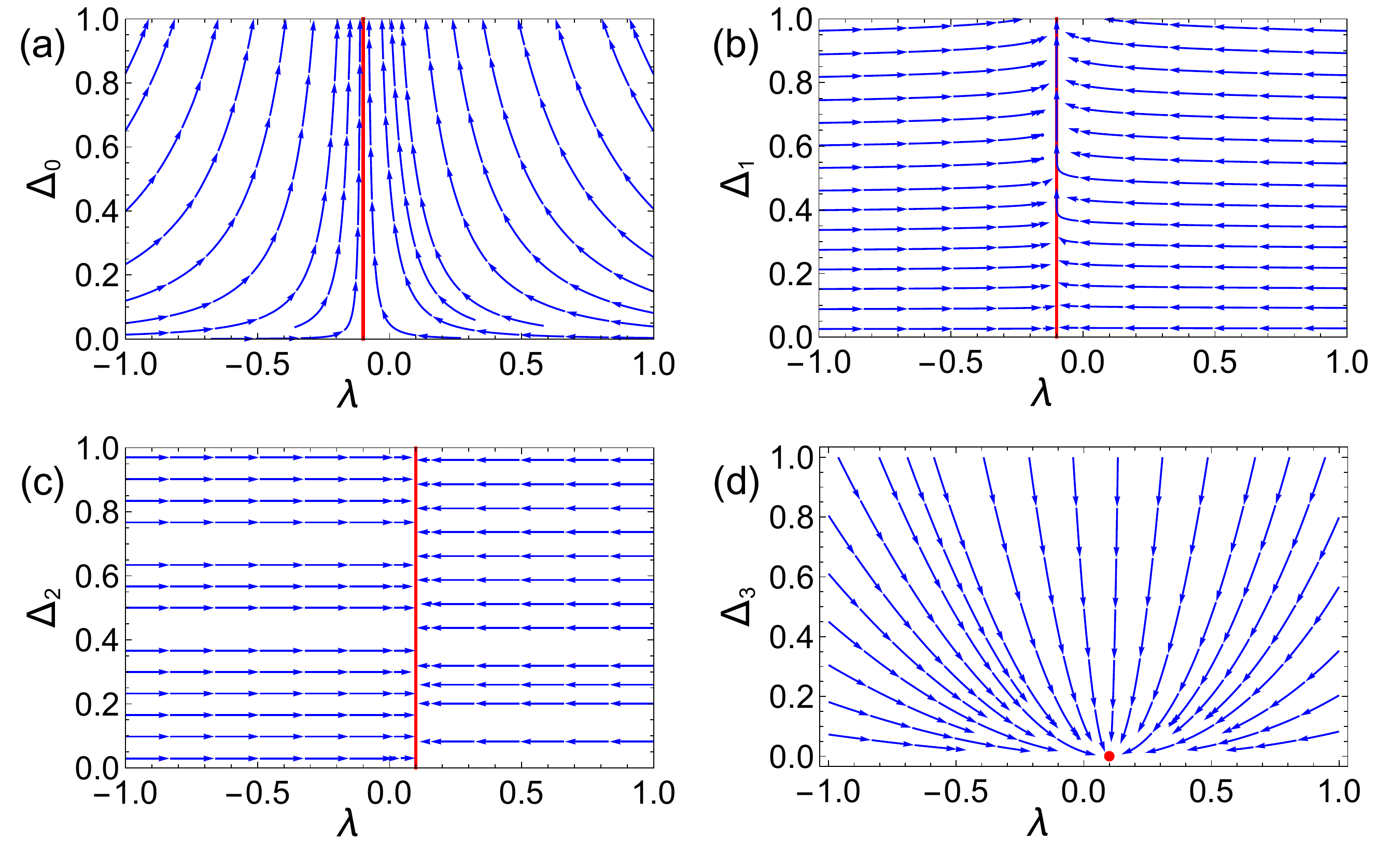}
\caption{ RG flow in the $\la-\De_i$ plane for single disorder exists. For all graphs, $t_0=0.1$ is used. } \label{FigFlDelac}
\end{figure}

We first pay attention to a single kind of disorder. Coexistence of different kinds of disorders is analyzed at last. According to \Eq{EqRGt}, there is no renormalization of the ordinary tilt $t$ if we set $\alpha=0$, which means that disorder does not influence the tilt solely. This is consistent with the previous result in Ref. \cite{Sikkenk17} that disorder does not renormalize the ordinary tilt for a 3D tilted Weyl semimetal. Due to the non-renormalization of $t$, in this section below, we fix $t\xk{\ell}=t_0$ with $t_0$ being a constant and $0<t_0\ll 1$. However, the observable tilt $t'$ is dependent on the ordinary tilt $t$ and the anomalous tilt $\la$. Its fate still needs to investigate by analyzing the disorder effect to $\la$.

\textit{RSP.} By setting $\De_1=\De_2=\De_3=0$, the left RG equations for $\De_0$, $\la$ and $v_x$ reduce to $d \De_0 /d\ell=2\De_0^2/\xk{1-t_0^2}$, $d \la/d\ell=-\De_0\xk{1-\la t_0}\xk{t_0+\la }/\xk{1-t_0^2}$, $d v_x / d \ell=-\xk{1-\la t_0}\De_0v_x/\xk{1-t_0^2}$. Solutions of these three differential equations are given by
\bea
\De_0\xk{\ell}&=&\frac{\De_0^0\xk{1-t_0^2}}{1-t_0^2-2\De_0^0\ell },\quad \la\xk{\ell} =\frac{t_0\zk{1-A_0\xk{\ell}} }{A_0\xk{\ell}+t_0^2},
\nn\\
v_x\xk{\ell}&=&v_x^0\zk{\frac{1+t_0^2}{A_0\xk{\ell}+t_0^2}} \label{Eqsolus0},
\eea
where $\De_0^0$, $v_x^0$ are the initial values of $\De_0$ and $v_x$ respectively, $A_0\xk{\ell}=\zk{\frac{\De_0\xk{\ell}}{\De_0^0}}^{\xk{1+t_0^2}/2}$ and the initial value of  $\la$ is assumed to be zero through our paper. Based on above solutions, we found
\bea
&&\De_0\xk{\ell}|_{\ell\rightarrow \ell_c} \rightarrow +\infty,\quad \la\xk{\ell}|_{\ell\rightarrow \ell_c} \rightarrow -t_0,
\nn\\ &&
v_x\xk{\ell}|_{\ell\rightarrow \ell_c} \rightarrow 0, \label{Eqfpvla0}
\eea
where $\ell_c=\xk{1-t_0^2}/2\De_0^0$, which corresponds to a constant energy scale as
$E_c=\La e^{-\ell_c}=\La e^{-\xk{1-t_0^2}/2\De_0^0}$.
Here, $\La$ is the UV cutoff which is determined by the bandwidth. Instead of the above analytical treatment, we have plotted the flow diagram in $\xk{\la,\De_0}$ plane in \Fig{FigFlDelac}(a), from which the flows of $\De_0$ and $\la$ read more directly. Both ways show that $\De_0\xk{\ell}$ flows to infinity at a constant energy scale and this divergence is interpreted as the phase transition from the DSM into a strongly disordered phase. We will show that this phase is just the well studied CDM phase which is identified as the existence of the same finite scattering rate for fermions in two orbitals \cite{Trescher17, Sikkenk17,Fradkin1, Fradkin2,Shindou09,Ominato14,Goswami11,Kobayashi14,Sbierski14,Roy14,
Syzranov15L,Syzranov15B,Pixley15}. Above result is suitable to untilted Dirac fermions if we set $t_0=0$. Therefore, tilt has not qualitatively changed the relevance of RSP, it is still marginally relevant. However, in the next subsection, we will reveal that tilt brings new physics for the CDM phase: the Dirac point will be replaced by a bulk nodal arc. Now, we explain the meanings of \Eq{Eqfpvla0} for $v_x$ and $\la$. It seems that $v_x$ should be zero in the CDM phase, however, we remind this is illusory because once the energy approach $E_c$, the RG breaks down. Actually, when the system enters into the CDM phase, the disorder strength becomes a large and finite value rather than the infinity \cite{Fradkin2,Shindou09,Ominato14,Sbierski14,
Syzranov15L,Syzranov15B}. Correspondingly, the fermion velocity keeps to be a nonzero constant and its square is proportional to the diffusion constant \cite{Altlandb,Colemanb} of the CDM phase. Incorporating the renormalization effect to $v_x$ before entering into the CDM phase, we conclude that $0<v_x^{\ast}<v_0$, where $v_x^{\ast}$ represents the constant velocity in the CDM phase. Above analysis also establishes for $\la$. In particular, in the CDM phase, $\la$ keep to be a constant as $\la_{\ast}$ and $-t_0<\la_{\ast}<0$. As a result, the observable tilt becomes
\bea
t'|_{\la=\la_{\ast}}=\frac{t_0-\la_{\ast}}{1-\la_{\ast}t_0} \in\xk{ t_0,\frac{2t_0}{1+t_0^2} }.
\eea
Noting that, for $t_0\ll 1$, $2t_0/\xk{1+t_0^2}\approx 2 t_0 \ll 1$. Therefore, RSP increases the tilt but the tilt is still a small finite value, which ensures the linearized model of \Eq{EqfrHam} to be well established. This result indicates that the RSP induced diffusive phase transition precedes the Lifshitz transition between type-I and type-II DSMs, which supports earlier conclusion obtained by numerically solving the SCBA equations in Ref. \cite{Trescher17}.

\textit{$x$-RVP.} With other disorders vanishing, the simplified RG equations for $\De_1$, $\la$ and $v_x$ are $d\De_1 /d\ell=2t_0^2\De_1^2/\xk{1-t_0^2}$, $d\la/d\ell=-\De_1\xk{1-\la t_0}\xk{t_0+\la }/\xk{1-t_0^2} $, $dv_x / d \ell=-\xk{1-\la t_0}\De_1v_x/\xk{1-t_0^2}$.
Expressions of $\De_1\xk{\ell}$, $\la\xk{\ell}$, and $v_x\xk{\ell}$ read as
\bea
\De_1\xk{\ell}&=&\frac{\De_1^0\xk{1-t_0^2}}{1-t_0^2-2t_0^2\De_1^0\ell },\quad \la\xk{\ell} =\frac{t_0\zk{1-A_1\xk{\ell}} }{A_1\xk{\ell}+t_0^2},
\nn\\
v_x\xk{\ell}&=&v_x^0\zk{\frac{1+t_0^2}{A_1\xk{\ell}+t_0^2}} \label{Eqsolus1},
\eea
where $\De_1^0$ is the initial value of $\De_1$ and $A_1\xk{\ell}=\zk{\frac{\De_1\xk{\ell}}{\De_1^0}}^{\xk{1+t_0^2}/2t_0^2}$. Therefore, $\De_1$ will unboundedly flow to infinity at the energy scale of $\ell'_c=\xk{1-t_0^2}/\xk{2\De_1^0t_0^2}$, which indicates that an arbitrarily weak $x$-RVP can induce the diffusive phase transition \cite{Fradkin1, Fradkin2,Shindou09,Ominato14,Goswami11,Kobayashi14,Sbierski14,Roy14,
Syzranov15L,Syzranov15B,Pixley15,Trescher17, Sikkenk17}. This behavior resembles RSP very much. The only difference is that tilt plays a vital role now. If we set $t_0=0$, there is $d \De_1 /d\ell=0$, we repeat the well-known result that RVP in untilted Dirac fermions is unrenormalized and this is guaranteed by the time-independent gauge transformation \cite{Ludwig94,Vafek08,Herbut08}. By comparing, we conclude that a nonzero tilt enhances the $x$-RVP to be a marginally relevant perturbation. Therefore, the RVP-induced diffusive phase transition is a purely new phenomenon which originates from the interplay between disorder and tilt. In the next section, we will show that a bulk nodal
arc appears in the CDM phase, which also originates from this interplay. As for the fates of $\la$, $t'$, and $v_x$ in this CDM phase induced by $x$-RVP, they are very similar to the cases for RSP, for which we do not state repeatedly here. The RG flow for $\xk{\la,\De_1}$ plane is shown in \Fig{FigFlDelac}(b), its similarity to \Fig{FigFlDelac}(a) is apparent.

\textit{$y$-RVP.} The reduced RG equations for  $\De_2$, $\la$ and $v_x$ are $d \De_2 /d\ell=0$, $d \la/d\ell=-\De_2\xk{1-\la t_0}\xk{\la-t_0 }/\xk{1-t_0^2} $, $d v_x / d \ell=-\xk{1-\la t_0}\De_2v_x/\xk{1-t_0^2}$. We see that the $y$-RVP is marginal at one-loop level. In fact, this result is established at any order of loop expansion due to the presence of a local time-independent gauge symmetry as
\bea
\psi_i \rightarrow \psi_i e^{i\xi(y)}, \quad A(\b x)\rightarrow A(\b x)+ \pa_y \xi(y).
\eea
The detailed proof of this un-renormalization was provided in Ref. \cite{PLZhao2018}. We remind that tilt along the $x$-direction can destroy this gauge invariance, for which reason, the $x$-RVP is renormalized. Because we have used an $x$-direction tilt at the beginning, the gauge symmetry dependent solely on $y$ coordinate still keeps intact. Suppose tilt changes to $\b t=\xk{t_x, t_y}\xk{t_x,t_y\neq 0}$, then the time-independent local gauge symmetry does not exist. A trivial check tells us that the $y$-RVP becomes a marginally relevant coupling, which is just what happened for $x$-RVP. Therefore, we have figured out that the tilt enhances the relevance of the $x$-RVP
is attributed to its ability to break the time-independent gauge symmetry \cite{Ludwig94,Vafek08,Herbut08} existing along the $x$-direction. For our present consideration, the $y$-direction is not tilted, as a result, the $y$-RVP is unrenormalized, and we take $\De_2\xk{\ell}=\De_2^0>0$. Substituting this value to the RG equation for $\la$,
the solution for $\la\xk{\ell}$ is obtained to be
\bea
\la\xk{\ell}=\frac{t_0\xk{1-e^{-\De_{2}^0\ell} }}{1-t_0^2e^{-\De_{2}^0\ell}},\quad v_x\xk{\ell}=\frac{v_x^0\xk{1-t_0^2}}{e^{\De_{2}^0\ell}-t_0^2}.
\eea
Based on these results, we found $\la\xk{\ell}\big|_{\ell\rightarrow \infty}=t_0$ and $v_x\xk{\ell}\big|_{\ell\rightarrow \infty} \sim e^{-\De_{2}^0\ell}\rightarrow 0$, which means that the stable fixed point for $\la$ is $\la_{\ast}=t_0$ and $v_x$ flows to zero exponentially. The flow diagram for $\xk{\la,\De_2}$ is presented in \Fig{FigFlDelac}(c), from which the marginality of $\De_2$ and the stability of $\la_{\ast}=t_0$ is clearly shown. For $\la_{\ast}=t_0$, the observable tilt turns to be
\bea
t'\big|_{\la_{\ast}=t_0}=0.
\eea
Therefore, the $y$-RVP reduces the observable tilt and stabilizes type-I Dirac fermions, which effect is different from RSP and $x$-RVP.

\textit{RM.} Similarly, vanishing of other types of disorders yields
the following flow equation $d \De_3 /d\ell=-2\De_3^2$, $d \la/d\ell=-\De_3\xk{1-\la t_0}\xk{\la-t_0 }/\xk{1-t_0^2} $, $d v_x / d \ell=-\xk{1-\la t_0}\De_3v_x/\xk{1-t_0^2}$. The analytical solutions are obtained by solving these two coupled differential equations, which take the form
\bea
\De_3\xk{\ell}&=&\frac{\De_3^0}{1+2\De_3^0\ell },\quad \la\xk{\ell}=\frac{t_0\xk{\sqrt{1+2\De_3^0\ell}-1} }{\sqrt{1+2\De_3^0\ell}-t_0^2   },
\nn\\
v_x\xk{\ell}&=&\frac{v_x^0\xk{1-t_0^2} }{\sqrt{1+2\De_3^0\ell}-t_0^2   } .
\eea
Therefore, $\De_3\xk{\ell}|_{\ell \rightarrow \infty} \rightarrow 0$, $\la\xk{\ell}|_{\ell \rightarrow \infty} \rightarrow t_0$, and $v_x\xk{\ell}|_{\ell \rightarrow \infty} \rightarrow 0$, which indicates the stable fixed point in the $\la-\De_3$ plane is  $\xk{\la_{\ast},\De_3^{\ast}}=\xk{t_0, 0}$. The RG flow plotted in \Fig{FigFlDelac}(d) also confirms this. At this stable fixed point, the observable tilt is also decreased and the type-I Dirac fermions are also stable.

\textit{Coexist.} When different kinds of disorder coexist,
\Eqs{EqRGDe0}--(\ref{EqRGDe3}) tell us that there is an
interesting correlation among different disorders:
if any two types of disorder coexist,
the other two can be dynamically generated and all of the four disorder
parameters will coexist. Therefore, we need to analyze the fixed point of
\Eqs{EqRGDe0}--(\ref{EqRGDe3}). There is only one constant fixed point
as $\xk{\De_0^{\ast},\De_1^{\ast},\De_2^{\ast},\De_3^{\ast}}=\xk{0,0,0,0}$.
However, this fixed point is not stable, all disorders flow to an uncontrolled
strong-coupling regime, which indicates the appearance of diffusive phase
transition \cite{Fradkin1, Fradkin2,Shindou09,Ominato14,Goswami11,
Kobayashi14,Sbierski14,Roy14,Syzranov15L,Syzranov15B,Pixley15,Trescher17, Sikkenk17}
again. Besides, this result is independent of Coulomb interaction, for
which we will no longer give a detailed analysis of coexistence of
Coulomb interaction with all disorders in the next section.

\textit{DOS and specific heat.} Having determined the fate of disorder couplings and tilt in low energy, we now focus on disorder influence to DOS and specific heat. With details shown in Appendix \ref{Sec:Apenob}, we just summarize the results in Table~\ref{TBDOS}. From which we see that the relevant RSP or $x$-RVP generates a nonzero zero-energy DOS in the CDM phase, the marginal $y$-RVP supplies a power-law enhancement to the low-energy DOS and specific heat, and the marginally irrelevant RM provides a logarithmic enhancement to the low-energy DOS and specific heat. In addition, a nonzero tilt increases the zero-energy DOS generated by RSP or $x$-RVP and the low-energy DOS provided by $y$-RVP.

\begin{table}[htbp]
\caption{DOS and specific heat without Coulomb interaction. Here, Dis. is the abbreviation for disorder, $f\xk{t_0}=\xk{1+t_0^2}/\xk{1-t_0^2}^{3/2}$ and $h\xk{t_0}=\xk{1+t_0^2}/\xk{1-t_0^2}$. Besides, $\ga_0$ and $\ga_0'$ represents fermion scattering rate induced by RSP and $x$-RVP respectively.\label{TBDOS}}
\vspace{-0.3cm}
\begin{center}
\begin{tabular}{|c|c|c|}
\hline\hline  Dis.  &  DOS  &  Specific heat
\\
\hline  None  &
$\rho(\omega)\sim\omega$  &  $C_{v}(T)\sim
T^{2}$
\\
\hline RSP  &  $\rho(0)\sim f\xk{t_0}\ga_0\ln\xk{\frac{\La}{\ga_0}} $
 &  $C_{v}(T)\sim  \rho(0)T$
\\
\hline $x$-RVP  &  $\rho(0)\sim f\xk{t_0} \ga'_{0}\ln\xk{\frac{\La}{\ga'_0}}$
 &  $C_{v}(T)\sim \rho(0) T$
\\
\hline  $y$-RVP  &
$\rho(\om)\sim\om^{1-2 h\xk{t_0}\De_{2}^0 }$   &
$C_{v}(T)\sim T^{2-2 h\xk{t_0} \De_{2}^0}$
\\
\hline  RM  &
$\rho(\om)\sim \om \ln \om$   &
$C_{v}(T)\sim T^2 \ln T$
\\
\hline\hline
\end{tabular}
\end{center}
\end{table}

\subsection{Bulk nodal arc \label{Sec:Bna} }

In the previous subsection, we have found that RSP and $x$-RVP both drive the system into a strongly disordered phase. Now, we employ the SCBA to study the properties of this phase. We will show that this strongly disordered phase corresponds to the CDM phase and the Dirac point is replaced by a bulk nodal arc in this phase.

In general, the zero-energy self-energy is proposed to be $\Si\xk{0}=\Si_0\s_0+\Si_1\s_1$. Using this ansatz, the self-consistent equation for self-energy is given by
\bea
\Sigma_{j}(0) =- \kappa_j \int'\frac{d^2k}{(2\pi)^2} \s_j \frac{1}{H_0\xk{\b k}+\Sigma_{j}(0) }\s_j \label{EqSCBA},
\eea
where $j=0,1$ to label RSP and $x$-RVP induced self-energy, respectively, $\int'$ represents an integration with the UV-cutoff $\La$. Finishing this integration yields
\bea
\Sigma_{1j}&=&-t\Si_{0j},\label{EqSi1}\\
\Si_{0j}&=&\pm i\frac{\La}{2} \exp\zk{\frac{-\xk{1-t^2} }{2\De_j}} \equiv \pm i\ga_{0j}\label{EqSi0},
\eea
where the $\ga_{0j}$ refers to the finite scattering rate for fermions. As a result, the retarded self-energy at zero-energy takes the form:
\bea
\Si_j^R\xk{0}=-i\ga_{0j}\xk{\s_0-t\s_1}\label{Eqreserp}.
\eea
Therefore, the scattering rates for fermions in $\s_z=+1$ and $-1$ orbitals are exactly the same. As a result, the strongly disordered phases for RSP and $x$-RVP are both well-defined CDM phases. However, for the model of tilted Dirac fermions considered in Refs. \cite{Papaj18,PLZhao2018}, the scattering rates for fermions in two orbitals are different and the corresponding strongly disordered phase is distinct from the traditional CDM phase and regarded as a new strongly disordered phase.

\begin{figure}[htbp]
\centering
\includegraphics[width=3.0in]{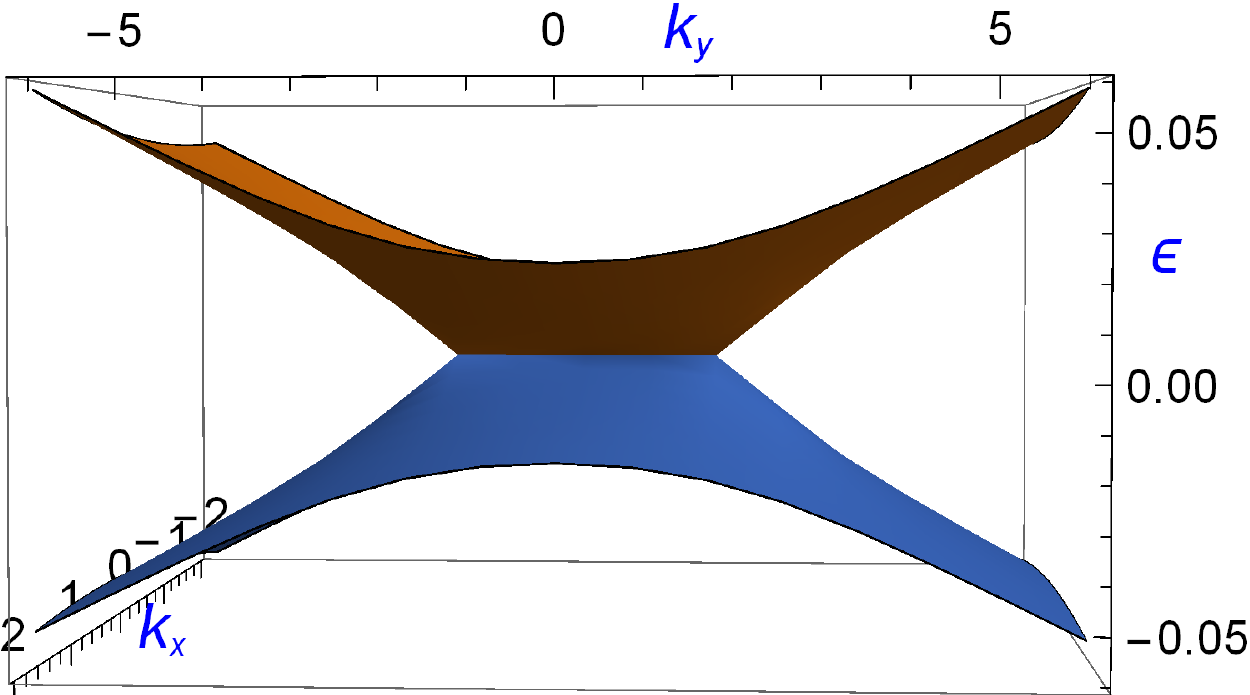}
\caption{Quasiparticle energy dispersion of \Eq{EqComdispRP} for tilted Dirac fermions in CDM phase. Here, the momentum and energy are
dimensionless and $t_0=0.1$, $\la=-0.09$, $v_x=v_y=0.01$, $\ga_{0j}=2$. } \label{Fignodalarc}
\end{figure}

After including the above retarded self-energy, the energy spectrum of the quasiparticles is found by calculating the poles of the disorder-averaged retarded Green's function as
\bea
\Det{\xk{G_0^R\xk{E,\b k}}^{-1}-\Si_j^R}=0.
\eea
Concretely,
\begin{widetext}
\bea
\Det{E\xk{\s_0+\la \s_1} - v_x t k_x\s_0 -v_x k_x \s_1-v_y k_y \s_2 +i\ga_{0j}\xk{\s_0-t\s_1}}=0.
\eea
Solution of $E$ reads as
\bea
E_{\pm}\xk{\b k}&=&\frac{1}{1-\lambda ^2}\Bigg\{ k_x v_x\xk{t-\la}-i \ga_{0j}\xk{1+\la t}
\pm \sqrt{\left[(1-\lambda  t) v_x k_x -i \ga_{0j} (\lambda +t)\right]{}^2+\left(1-\lambda ^2\right) v_y^2k_y^2} \Bigg\}.
\eea
Here, we remind that all the parameters are the renormalized ones in the CDM phase but not the bare ones. Absolutely, $E_{\pm}\xk{\b k}$ is a complex value and its real part corresponds to the quasiparticle energy-momentum relation \cite{Shen2018, Kozii2017,
Papaj18}, which can be written as
\bea
E_{\pm}^{\rm R}=\frac{1}{1-\lambda ^2}\zk{k_x v_x\xk{t-\la} \pm \frac{1}{\sqrt{2}}\sqrt{\sqrt{{\cal
E}_0^2 + 4\ga_{0j}^2 \xk{\lambda +t}^2\xk{1-\lambda  t}^2 v_x^2 k_x^2} + {\cal E}_0 }}, \label{EqComdispRP}
\eea
\end{widetext}
where ${\cal E}_0=\xk{1-\lambda  t}^2 v_x^2 k_x^2+\xk{1-\lambda ^2} v_y^2k_y^2-\ga_{0j}^2 \xk{\lambda +t}^2$. By requiring $E_{\pm}^{\rm R}=0$, we obtain a line along the $y$-direction as
\bea
k_x=0, \quad  -\frac{\ga_{0j}\xk{\lambda +t} }{v_y \sqrt{1-\lambda ^2}} \leq k_y \leq \frac{\ga_{0j}\xk{\lambda +t} }{v_y  \sqrt{1-\lambda ^2}}\label{Eqbulkarc}.
\eea
Therefore, instead of a Dirac point, the Fermi surface transforms into a bulk nodal arc described by \Eq{Eqbulkarc}. To show this bulk nodal arc more clearly, we plotted the energy-momentum relation in \Fig{Fignodalarc}. Along with this bulk nodal arc, conduction and valence bands are degenerate naturally. The two end points of this arc are exceptional points, which are treated as a split of the Dirac point \cite{Heiss2012,Shen2018}. Due to the existence of two end points, this bulk nodal arc is different from the two open lines for the Fermi surface of type-II Dirac fermions \cite{Goerbig08}, which indicates that appearance of this bulk nodal arc is nothing to do with the Lifshitz transition \cite{Goerbig08,Isobe16}. Actually, we have shown that the effective tilt keeps being a small value in the CDM phase, which excludes the appearance of the Lifshitz transition.  However, in Refs. \cite{Papaj18,PLZhao2018}, the appearance of the bulk nodal arc always accompanies the Lifshitz transition, which means that the effective tilt $t \rightarrow 1$ with the system entering the strongly disordered phase. The difference mentioned above is related to the nature of the strongly disordered phase. As we explained above, the strongly disordered phase in Refs. \cite{Papaj18,PLZhao2018} is distinct from the CDM phase and for the new strongly disordered phase in Refs. \cite{Papaj18,PLZhao2018}, the tilt is dynamically generated and approaches to the unit even if it is absent at the beginning. For the model considered in this work, the strongly disordered phase is a well-defined CDM phase and the tilt cannot be dynamically generated. Furthermore, the origin of the bulk nodal arc in this work is also different from the one in Refs. \cite{Papaj18,PLZhao2018}. In Refs. \cite{Papaj18,PLZhao2018}, a bulk nodal arc
is produced due to the appearance of different scattering rates for fermions in different orbitals, which is a natural result because fermions in two different orbitals are not related by any symmetry. Actually, the Brillouin zone for the model considered in Refs. \cite{Papaj18,PLZhao2018} respects no external symmetry at all. However, for the model considered in this work, the ``particle-hole" symmetry still holds and the scattering rates for two orbitals are the same. The emergence of the bulk nodal arc is attributed to the term of $i\ga_{0j} t_0\s_1$ in \Eq{Eqreserp}, which does not commute with $H_0$ but respects the ``particle-hole" symmetry. Suppose that no tilt exists at the beginning or no finite scattering rate is generated, the above term will vanish and no bulk nodal arc appears. Therefore, the appearance of the bulk nodal arc in this work is just a result coming from the interplay of tilt and certain type of disorder (RSP or $x$-RVP), which guarantees the appearance of $t_0$ and $\ga_{0j}$ at the same time and ensures the existence of $i\ga_{0j} t_0\s_1$. Although the model studied in this work is totally different from the model in Refs. \cite{Papaj18,PLZhao2018}, it can be seen as a 2D resemblance of the 3D tilted Weyl semimetals considered in Ref. \cite{Zyuzin18}, and the bulk nodal arc found here can be viewed as a 1D resemblance of the 2D circle flat band obtained in Ref. \cite{Zyuzin18}.

\section{Interplay with Coulomb interaction \label{Sec:IwCi} }

\begin{figure}[htbp]
\center
\includegraphics[width=3.3in]{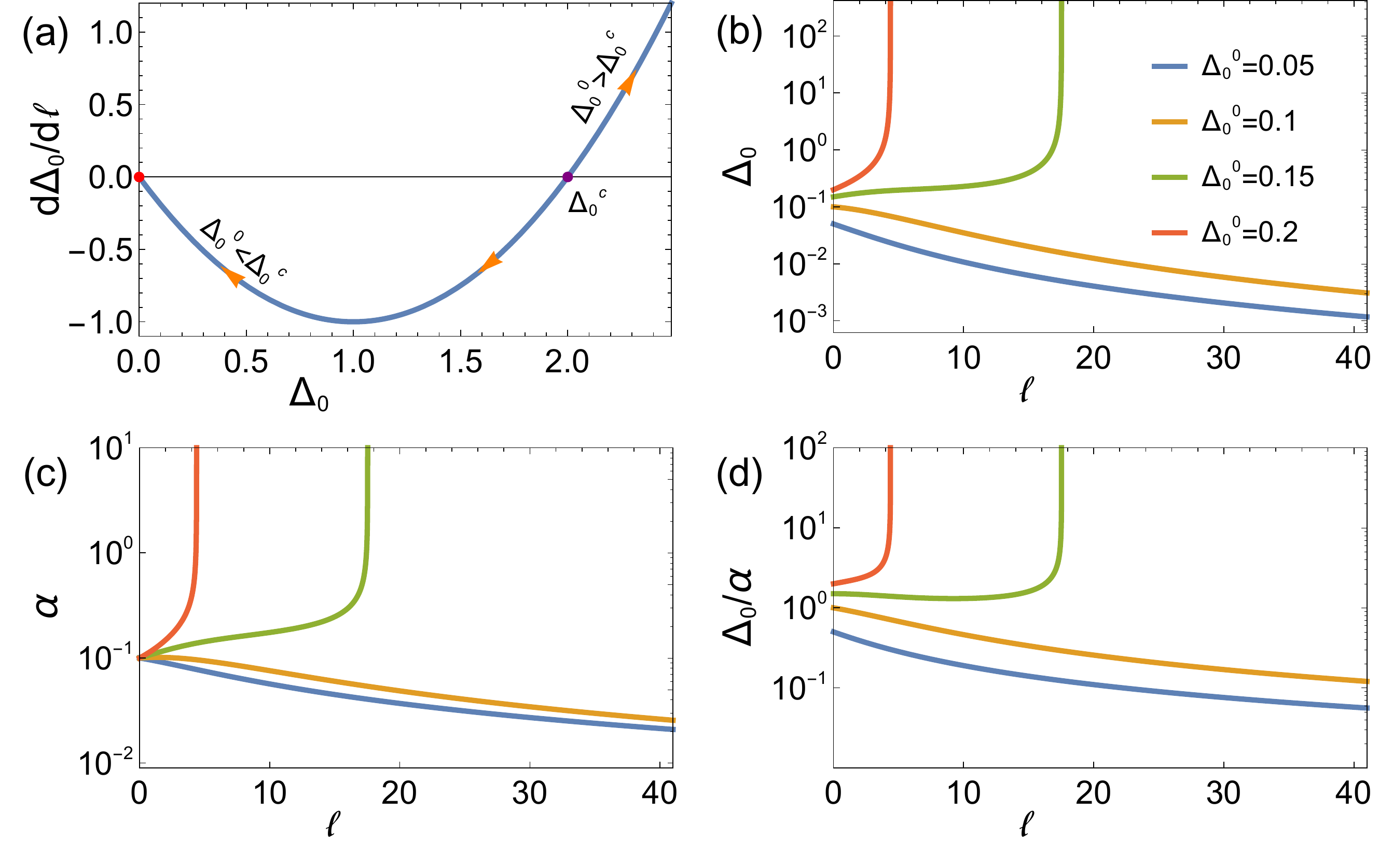}
\caption{(a) Shape for the dependence of $\frac{d\De_0}{d\ell}$ on $\De_0$, (b)--(d) flow of $\De_0$, $\alpha$, and $\De_0/\alpha$. All diagrams are plotted in the presence of RSP and Coulomb interaction. In (a), $\De_0^c$ is a constant given by \Eq{EqcriDe0} with determined initial values for $\alpha$, $t$, $\la$, and $\eta$. In (b)--(d), $t_0=0.5$, $\la_0=0$, $v_x^0=0.5$, $\eta_0=0.5$, $\alpha_0=0.1$, as a result $\De_0^c\approx 0.117$. The labels of colored lines for (c) and (d) are the same as (b).} \label{De0Coall}
\end{figure}

We now consider the interplay between single disorder
and Coulomb interaction. The rest of this subsection is organized similarly as the previous one.

\begin{figure}[htbp]
\center
\includegraphics[width=3.3in]{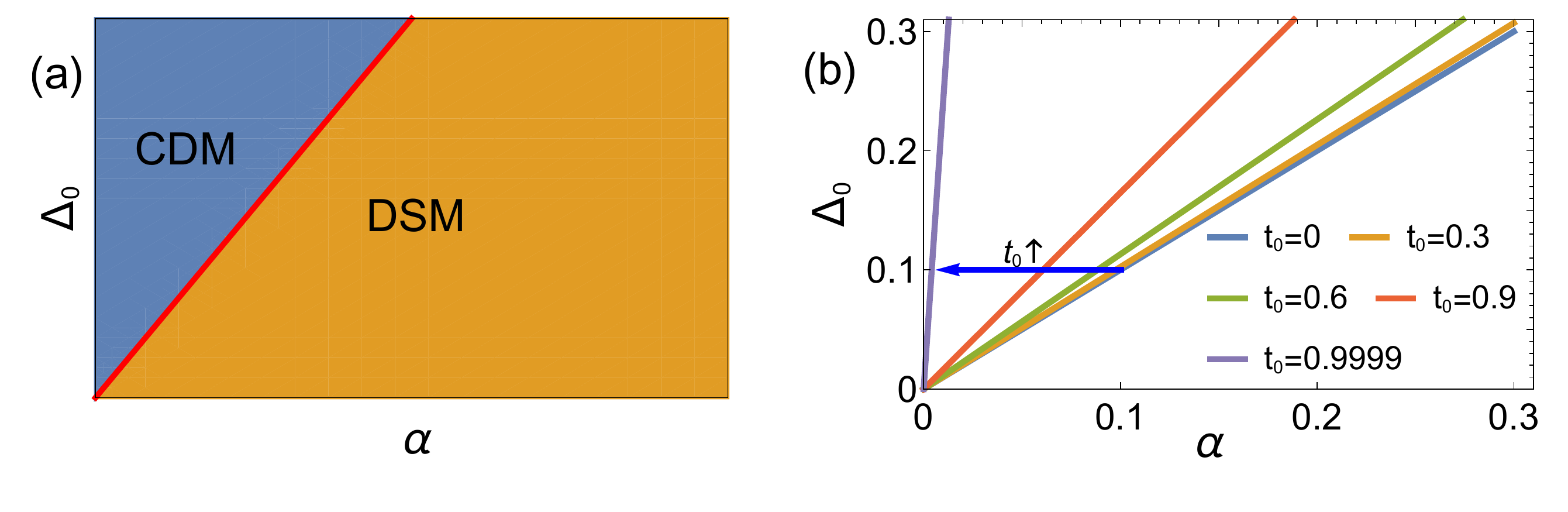}
\caption{(a) Phase diagrams of the tilted DSM in the presence of RSP
and Coulomb interaction. (b) Influence of $t_0$ on the boundary of two phases. For two graphs, $\la_0=0$, $\eta_0=1$ is used. } \label{Figphadia}
\end{figure}

{\it RSP.} Again, by setting $\De_1=\De_2=\De_3=0$ and $\alpha \neq 0$, we obtain the required RG equations to perform the further analysis. According to simplified \Eq{EqRGDe0}, we plot the sketch for the dependence of $d\De_0/d\ell$ on $\De_0$ in \Fig{De0Coall}(a), which clearly shows that a critical RSP strength appears due to the interplay of Coulomb interaction and RSP. The critical RSP strength can be written as
\bea
\De_0^c=\frac{\alpha\xk{1-t^2} }{2}\zk{f_x\xk{\frac{2\la^2}{1-\la t} + \frac{1-\la t}{1-t^2}  }+ f_y}.\label{EqcriDe0}
\eea
Here, $\alpha$, $t$, $\la$, and $\eta$ are all referred to its initial value, as a result, $\De_0^c$ is just a constant. As exhibited in \Fig{De0Coall}(a), if $\De_0^0$ is larger than $\De_0^c$, $\De_0$ flows to the strong disorder regime. Suppose $\De_0^0<\De_0^c$, then $\De_0$ goes to zero in the low-energy limit. We also plotted the numerical solutions for $\De_0\xk{\ell}$, $\alpha\xk{\ell}$, and $\De_0\xk{\ell}/\alpha\xk{\ell}$ in \Figs{De0Coall}(b)--\ref{De0Coall}(d) respectively. According to \Figs{De0Coall}(c) and \ref{De0Coall}(d), we see that $\alpha$ and $\De_0/\alpha$ both flow to zero if $\De_0$ plotted in \Fig{De0Coall}(b) goes to zero. This means that RSP and Coulomb interaction are both very weak coupling and the Coulomb interaction plays a more overwhelming role than RSP. The dominated but weak Coulomb interaction guarantees the system to stay in the DSM phase. However, as depicted in \Figs{De0Coall}(c) and \ref{De0Coall}(d), $\alpha$ and $\De_0/\alpha$ follow $\De_0$ to go to infinity. Therefore, RSP and Coulomb interaction both flow to strong coupling region and the system is mainly controlled by the disorder. As a consequence, the DSM phase is not stable and the diffusive phase transition from DSM to CDM happens. We notice that the above results for the interplay between the RSP and Coulomb interaction
is very similar to that of 2D DSM with untilted Dirac fermion \cite{Ye98,Ye99,Stauber05,WangLiu14} as well as the multi-Weyl semimetals \cite{JRWang17}.
Based on the above analysis, we plot the phase diagram in the plane of $(\alpha,\De_0)$ in \Fig{Figphadia}(b). Obviously, the DSM and CDM phases are separated by a critical line, which is determined by \Eq{EqcriDe0}
with certain initial values for $t$, $\la$, and $\eta$. By considering that the anomalous tilt $\la$ does not exist in the clean system, we just set $\la_0=0$. Besides, for the clean system, the velocity anisotropy is irrelevant in the low energy \cite{ZMHuang17,YWLee18}, for which we choose $\eta_0=1$. Therefore, we mainly concern the influence of ordinary tilt to the critical line. We have plotted several critical lines for different values of $t_0$ in \Fig{Figphadia}(b), from which we see that increase of $t_0$ enhances the critical value of $\De_0$ for a given $\alpha$. Therefore, the ordinary tilt is helpful to stabilize the DM phase.

\begin{figure}[htbp]
\center
\includegraphics[width=3.3in]{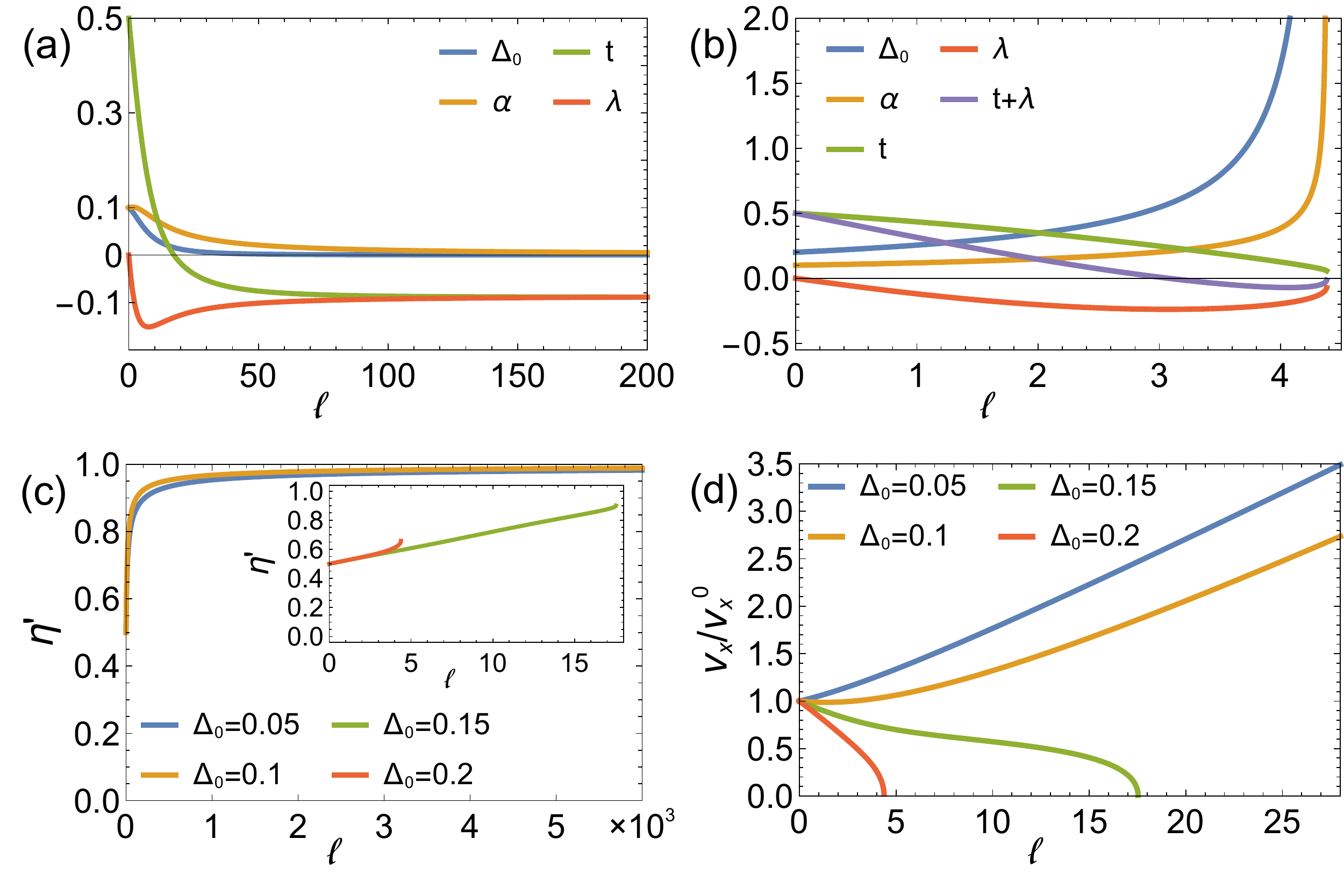}
\caption{(a) Flow of $t$ and $\la$ for $\De_0^0=0.1<\De_0^c$. (b) Flow of $t$, $\la$, and $t+\la$ when $\De_0^0=0.3>\De_0^c$. (c), (d) Flow of $\eta'$ and $v_x$ with fixed $\alpha_0$ and different initial value of $\De_0$. Here, the values of $\De_0$ labeled in the graphs represents their initial value. For all graphs, $t_0=0.5$, $\la_0=0$, $v_x^0=0.5$, $\eta_0=0.5$, $\alpha_0=0.1$, as a result $\De_0^c\approx 0.117$. } \label{FigDe01}
\end{figure}
We then focus on the fate of observable tilt in two phases. The running for $\la\xk{\ell}$, $t\xk{\ell}$ with $\ell$ are plotted in \Fig{FigDe01} by solving the simplified RG equations numerically. Figure \ref{FigDe01}(a) shows that $\De_0$ and $\alpha$ vanish in low-energy limit, which represents the DSM phase. At the same times, $\la_{\ast}=t_{\ast}=C_S$ takes place when $\ell \rightarrow \infty$. For the CDM phase, from \Fig{FigDe01}(b), we found that with $\De_0$ approaching to infinity at a constant energy scale $\ell_c^{''}$, $\la_{\ast}=-t_{\ast}=C_D$ happens. Here, $C_S$ and $C_D$ are two constants in $\xk{-1,1}$. The above behaviors for $\la$, $t$ in two phases can be easily understood as below. For the DSM phase, Coulomb interaction dominates the low-energy behavior of the system, and the relationship of $t$ and $\la$ is mainly controlled by \Eq{EqRGt}, which tells us $t$ will flow to $\la$. Besides, $\De_0^{\ast}=0$ ceases the flow of $d\la/d\ell$ and permits a constant value for $\la$. As a result, we have $\la_{\ast}=t_{\ast}=C_S$ in the low-energy limit. For the CDM phase which is controlled by disorder, the relationship of $t$ and $\la$ is determined by \Eq{EqRGla}, which gives $\la_{\ast}=-t_{\ast}=C_D$ when $\ell=\ell_c^{''}$. However, according to our argument in Sec.\ref{Sec:Fp}, $\De_0$ is not really divergent and $C_D= -t_{\ast}<\la_{\ast}< 0$ in the CDM phase. Based on the fate of $\la$, $t$ in two phases, we find $t'_{\ast}=0$ in the DSM phase and it becomes a constant belonging to $\xk{C_D,2C_D/\xk{1+C_D^2}}$ in the CDM phase. Therefore, we conclude that the Lifshitz transition between two types of Dirac fermions does not happen in the presence of Coulomb interaction and RSP, no matter which phase the system corresponds to.

We now analyze the behaviors of $v_x$ and velocity anisotropy. For the DSM phase, $\la_{\ast}=t_{\ast}$ in low energy. We employ this relation to simplify the RG equations for velocity anisotropy to be
\bea
\frac{d \eta}{d\ell}&=& \frac{\eta\alpha}{\pi}  \int_0^{2\pi}\frac{\xk{\sin^2\th-\cos^2\th}  d\th}{ \sqrt{\cos^2\th+\xk{\eta/\sqrt{1-\la_{\ast}^2} }^2\sin^2\th}    }. \label{Eqetasm}
\eea
By defining $\eta'=v_y'/v_x'$, we found $\eta'\big|_{t_{\ast}=\la_{\ast}}=\eta/\sqrt{1-\la_{\ast}^2}$. Then we perform the integration for $\th$, the result is an analytic function for $\eta'$, which is labeled as $h\xk{\eta'}$. Based on $h\xk{\eta'}$, we found that there exist two fixed points for $\eta$. The unstable one is $\eta=0$ and another stable one is $\eta_{\ast}=\sqrt{1-\la_{\ast}^2}$, exactly, $\eta'=1$. To show this more directly, we plotted the numerical solutions for $\eta'\xk{\ell}$ in \Fig{FigDe01}(c). The
main figure presents the result for the DSM phase, from which we observe that $\eta'\xk{\ell} \rightarrow 1$ with $\ell\rightarrow \infty$. The inset represents $\eta'\xk{\ell}$ for the CDM phase, which tells us that $\eta'\xk{\ell}$ ceases to increase at $\ell_c{''}$ and becomes a constant smaller than the unit. In addition, solutions for $v_x\xk{\ell}$ are also plotted in \Fig{FigDe01}(d). From \Fig{FigDe01}(d), we see that $v_x\xk{\ell}$ increases linearly with sufficiently large $\ell$, which indicates a logarithmic enhancement of $v_x$ in low energy of the DSM phase. Such a velocity enhancement has been early predicted in graphene \cite{Gonzalez94} and is experimentally observed by Shubnikov\textendash de Haas oscillations \cite{Elias11,Kotov12}. For tilted Dirac fermions system, experiment performed by site-selective
nuclear magnetic resonance in $\alpha$-($\mathrm{BEDT}$-$\mathrm{TTF}$)$_{2}\mathrm{I}_{3}$
\cite{Hirata16} also confirms this velocity enhancement. Previous works \cite{Isobe12,Hirata16, ZMHuang17, YWLee18} demonstrate that this velocity enhancement attributes to the Coulomb interaction. Therefore, appearance of this velocity enhancement confirms our previous conclusion that Coulomb interaction dominates the DSM phase in low energy. For the CDM phase, the system is controlled by disorder which reduces the velocities exponentially in the RG scheme. As shown in \Fig{FigDe01}(d), $v_x$ flows to zero at $\ell_c^{''}$. However, we remind here that instead of zero, $v_x$ is a constant in the CDM phase.

At last, we would like to highlight the influence of Coulomb interaction to the RSP-induced bulk nodal arc which was found in Sec. \ref{Sec:Bna}. We have demonstrated that the CDM phase appears only when $\De_0^0>\De_0^c$. Once this relation does not meet, the system stays in a DSM phase without scattering rate generating. Therefore, the bulk nodal arc appears only when $\De_0^0>\De_0^c$. Remind that an arbitrarily weak RSP can induce a bulk nodal arc if there is no Coulomb interaction. By comparison, we conclude that Coulomb interaction is a detrimental perturbation to RSP induced bulk nodal arc.

\begin{figure}[htbp]
\center
\includegraphics[width=3.3in]{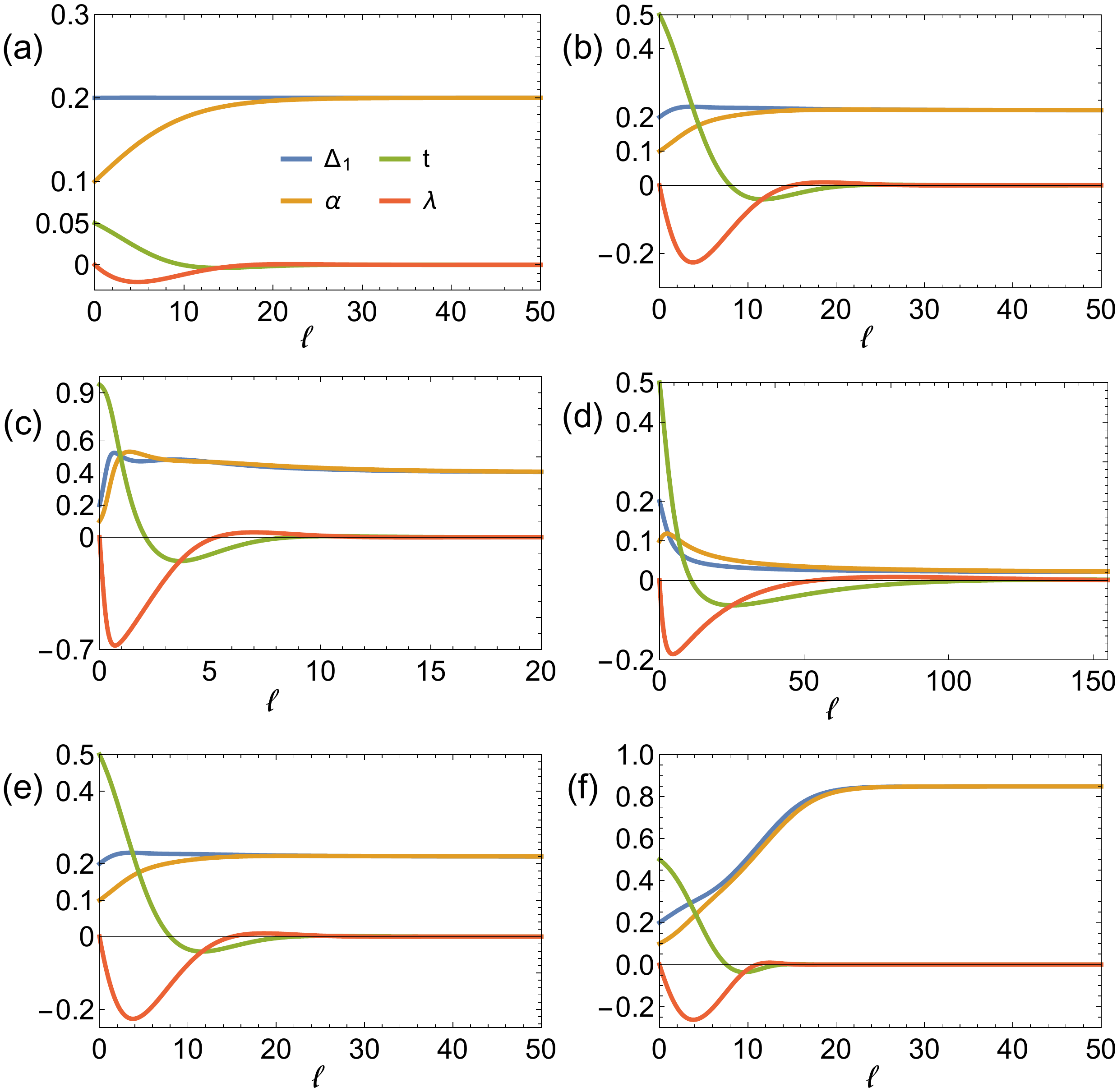}
\caption{Flow of $\De_1$, $\alpha$, $t$, and $\la$. The labels of colored lines for (b)--(f) are the same as (a). For all diagrams: $\la_0=0$, $v_x^0=0.5$, $\De_1=0.2$, $\alpha_0=0.1$. (a)--(c) $\eta_0=1$ and $t_0=0.05, 0.5,0.95$, respectively. (d)--(f) $t_0=0.5$ and $\eta_0=0.1, 1, 3.5$, respectively.  } \label{FigDe11}
\end{figure}

\begin{figure}[htbp]
\center
\includegraphics[width=3.3in]{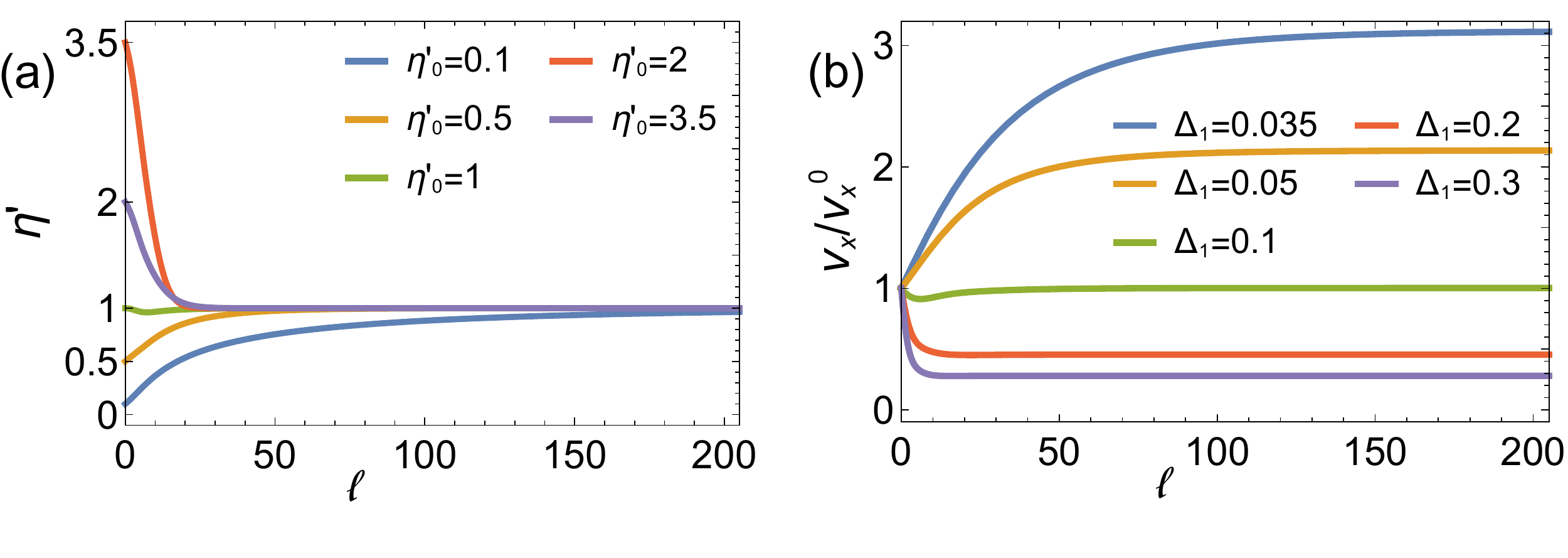}
\vspace{-0.5cm}
\caption{(a), (b) Flow of $\eta$ and $v_x$, respectively. For both graphs, $\la_0=0$, $t_0=0.5$, $v_x^0=0.5$, $\alpha_0=0.1$. In (a) $\De_1^0=0.2$, with $\eta_0$ changing. In (b) $\eta_0=1$, with $\De_1^0$ varying.} \label{FigDe12}
\end{figure}

{\it $x$-RVP.} By setting other disorder couplings to be zero and $\alpha \neq 0$, the simplified RG equations for the coexistence of $x$-RVP and Coulomb interaction are obtained. According to these RG equations, we found an interesting fixed point in the multi-parameter plane, which is given by $\xk{\De_1^{\ast}, \alpha_{\ast}, t_{\ast}, \la_{\ast}, \eta_{\ast}}=\xk{C_1, C_1, 0, 0, 1 }$. Here, $C_1$ is a nonzero constant and its concrete value is dependent on the initial value of all parameters in the reduced RG equations. This multi-parameters fixed point is easily understood. At the beginning, we suppose $\De_1^{\ast}$ and $\alpha_{\ast}$ can not be zero. A nonzero $\De_1$ gives $\la_{\ast}=-t$ according to simplified \Eq{EqRGla}, and a nonzero $\alpha$ renders $t_{\ast}=\la$ by employing simplified \Eq{EqRGt}. As a result, $t_{\ast}=\la_{\ast}=0$. Based on previous analysis of effective velocity anisotropy in DSM phase for RSP, $t_{\ast}=\la_{\ast}=0$ generates $\eta_{\ast}=\eta'_{\ast}=1$. According to simplified \Eq{EqRGDe1}, $d\De_1/d\ell=0$ once $t_{\ast}=\la_{\ast}=0$ and $\eta_{\ast}=1$. As a result, $\De_1$ neither flows to strong coupling regime nor goes to zero, it stays to be a constant in the low energy limit. At the attractive fixed points of $\xk{\De_1^{\ast}, t_{\ast}, \la_{\ast}, \eta_{\ast}}=\xk{C_1, 0, 0, 1 }$, the simplified \Eq{EqRGDe1} tells us that the attractive fixed point for $\alpha$ is $\alpha_{\ast}=\De_1^{\ast}$. Therefore, the starting point is guaranteed as neither $\De_1^{\ast}$ nor $\alpha_{\ast}$ is zero. To show this multi-parameter fixed point more directly, we have plotted several diagrams for the flow of $\De_1\xk{\ell}, \alpha\xk{\ell}, t\xk{\ell}, \la\xk{\ell}$ in \Fig{FigDe11}. Figures \ref{FigDe11}(a)--\ref{FigDe11}(c) reveal the influence of $t_0$, and \Figs{FigDe11}(d)--\ref{FigDe11}(f) display the effects of $\eta_0$. These two parameters are the two most important factors that determined the flow of $\De_1\xk{\ell}$. As we can see from \Fig{FigDe11}, no matter how the parameters change, $\De_1, \alpha, t, \la$ always flow to the fixed point given above. However, for different $t_0$ and $\eta_0$, $C_1$ is not the same in general. The flow of $\eta\xk{\ell}$ is shown in \Fig{FigDe12}(a), according to which, we see that $\eta$ goes to unit certainly. We now consider the consequence generated by this fixed point. Firstly, $v_x$ flows to a nonzero constant in low energy, we have shown this in \Fig{FigDe12}(b). Figure \ref{FigDe12}(b) also demonstrates that $v_x$ is enhanced if $\De_1^0$ is smaller than $\alpha_0$, in contrast, it is reduced as $\De_1^0$ is larger than $\alpha_0$. Secondly, due to $t_{\ast}=\la_{\ast}=0$, we obtain $t'_{\ast}=0$, which implies that type-I Dirac fermions are also stable against the coexistence of Coulomb interaction and $x$-RVP.

Another remarkable conclusion is that the bulk nodal arc is absent now. The Coulomb interaction prevents $x$-RVP to induce the diffusive phase transition and there is no scattering rate produced. Therefore, the Dirac point recovers when $x$-RVP coexists with Coulomb interaction.

\begin{figure}[htbp]
\center
\includegraphics[width=3.3in]{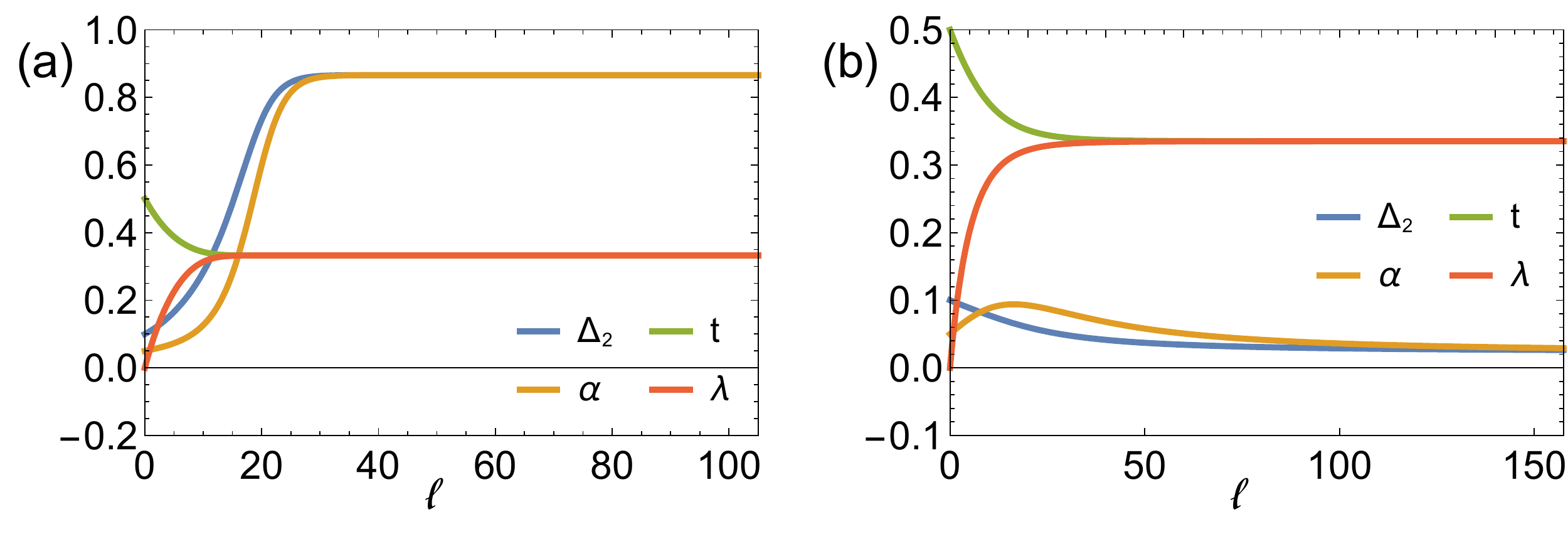}\vspace{-0.5cm}
\caption{Flow of $\De_2$, $\alpha$, $t$ and $\la$. For both diagrams: $\la_0=0$, $t_0=0.5$ $v_x^0=0.5$, $\De_2=0.1$, $\alpha_0=0.05$. In (a) $\eta_0=0.1$. In (b) $\eta_0=3.5$. } \label{FigDe2}
\end{figure}

{\it $y$-RVP.} The reduced RG equations for coexistence of $y$-RVP and Coulomb interaction can be obtained by setting $\De_0=\De_1=\De_3=0$. According to which, we found there exists another fixed point in the multi-parameter plane, which can be written as $\xk{\De_1^{\ast}, \alpha_{\ast}, t_{\ast}, \la_{\ast}, \eta_{\ast}}=\xk{C_2, C_2, C_t, C_t, \sqrt{1-C_t^2 }}$. Here, $C_2$ and $C_t$ are both nonzero constants. Still, we plotted diagrams for the flow of $\De_1\xk{\ell}, \alpha\xk{\ell}, t\xk{\ell}, \la\xk{\ell}$ in \Fig{FigDe2}. To simplify, we only consider the influence of $\eta_0$ and plotted diagrams for two different $\eta_0$. As shown in \Fig{FigDe2}, although the typical values of $C_2, C_t$ are varied with $\eta_0$, $\De_2^{\ast}=\alpha_{\ast}$ and $t_{\ast}= \la_{\ast}$ are always established. According to the analysis for $x$-RVP, $\De_2^{\ast}=\alpha_{\ast}$ renders a constant $v_x$ in the low energy limit and $t_{\ast}= \la_{\ast}$ guarantees $\eta'_{\ast}=1$, which implies $\eta=\sqrt{1-C_t^2}$. The flow of $\eta'\xk{\ell}$ and $v_x\xk{\ell}/v_x^0$ takes the similar form as they are shown in \Fig{FigDe12}, for which we do not draw them here. Besides, $t_{\ast}= \la_{\ast}$ also gives $t'_{\ast}=0$, which tells us type-I Dirac fermions are stable.

{\it RM.} The simplified RG equations for coexistence of RM and Coulomb interaction are obtained by requiring $\De_0=\De_1=\De_2=0$. Based on these simplified equations, we found a fixed point for the multi-parameter plane as $\xk{\De_3^{\ast}, \alpha_{\ast}, t_{\ast}, \la_{\ast}, \eta_{\ast}}=\xk{C_3, C_3, C_{\la}, C_{\la}, \sqrt{1-C_{\la}^2} }$, which is qualitative the same as the one obtained for the case of $y$-RVP. Therefore, the fate of $v_x$, $\eta'$, and $t'$ keep the same with previous results obtained for interplay between $y$-RVP and Coulomb interaction.

{\it DOS and specific heat.} In conclusion, the interplay of Coulomb interaction with RVP or RM generates identically stable fixed points for both interactions. At this stable fixed points, the effect tilt is irrelevant in low energy and the Dirac fermions stay to be type-I with velocity anisotropy vanishing. As a result, the corresponding system resembles graphene with the same interactions coexistence in low energy \cite{WangLiu14}. To show this similarity, we recalculate the DOS and specific heat. With details given in Appendix \ref{Sec:Apenob}, we display directly the low-energy asymptotic behavior for DOS and specific heat as
\bea
\rho\xk{\om} \sim \om, \quad C_v\xk{T} \sim T^2.\label{EqDOSshrfp}
\eea
Regardless of the prefactors, the DOS and specific heat share the same exponents with the free ones. If we introduce the dynamical exponent \cite{Hertz76, Lohneysen07, Herbutb}, which describes how the energy should be rescaled relative to the momenta, general scaling analysis shows that the
DOS should satisfy $\rho\xk{\om} \sim \om^{d/z-1}$ \cite{Fradkin2,Ludwig94} and the specific heat should meet $C_v\xk{T} \sim T^{d/z}$ \cite{Lohneysen07, Continentinob}. Therefore, \Eq{EqDOSshrfp} implies that $z=1$ when Coulomb interaction coexists with RVP or RM, which is consistent with the result studied for anisotropy graphene in Ref. \cite{WangLiu14} and previous general argument in Ref. \cite{Herbut01}. In addition, $\rho\xk{\om} \sim \om$ tells us that there is no anomalous dimension \cite{Khveshchenko2001,Franz2002,Gusynin2003,Herbut2009} generated for fermion field. Actually, Coulomb interaction and disorder produce anomalous dimensions separately, however, these two anomalous dimensions cancel each other in low-energy limit due to $\De_i^{\ast}=\alpha_{\ast}(i=1,2,3)$. The result that $v_x$ is a constant in the low-energy limit also confirms this cancellation.

\begin{table*}[htbp]
\caption{Summary of the main results in this work, including the flow behaviors of the coupling constants, effective tilt $t^{\prime}$, and the fermion velocity $v_x$, the possible phases due to the existence of certain interaction(s), and whether the bulk nodal arc exists. Here, CI is the abbreviation for Coulomb interaction and RSS stands for random stable state. \label{Summary}}
\vspace{-0.3cm}
\begin{center}
\begin{tabular}{|c|c|c|c|c|c|c|}
\hline\hline
Interaction(s)  & Coupling strength  &  $t^{\prime}=\frac{t-\la }{1-t\la}$  & $v_x$ & Phase & Bulk nodal arc
\\\hline
RSP  &  $\De_0\xk{\ell}|_{\ell\rightarrow \ell_c} \rightarrow +\infty$
 & $t^{\prime}|_{\ell\rightarrow \ell_c} \rightarrow \frac{2t_0}{1+t_0^2}$ & $v_x\xk{\ell}|_{\ell\rightarrow \ell_c} \rightarrow 0$& CDM & Yes
\\\hline
$x$-RVP  &  $\De_1\xk{\ell}|_{\ell\rightarrow \ell'_c} \rightarrow +\infty$
   & $t^{\prime}|_{\ell\rightarrow \ell'_c} \rightarrow \frac{2t_0}{1+t_0^2}$ & $v_x\xk{\ell}|_{\ell\rightarrow \ell'_c} \rightarrow 0$ &  CDM& Yes
\\ \hline
$y$-RVP  & $\De_2\xk{\ell}=\De_2^0$  & $t^{\prime}|_{\ell\rightarrow +\infty} \rightarrow 0$  & $v_x\xk{\ell}|_{\ell\rightarrow +\infty} \rightarrow 0$ &DSM & No
\\\hline
RM  & $\De_3\xk{\ell}|_{\ell\rightarrow +\infty} \rightarrow 0$   & $t^{\prime}|_{\ell\rightarrow +\infty} \rightarrow 0$  & $v_x\xk{\ell}|_{\ell\rightarrow +\infty} \rightarrow 0$ &DSM  & No
\\\hline
RSP$+$CI &$\De_0^0>\De_0^c: \De_0\xk{\ell}|_{\ell\rightarrow \ell''_c} \rightarrow +\infty $  & $t^{\prime}|_{\ell\rightarrow \ell''_c} \rightarrow \frac{2t_0}{1+t_0^2} $ & $v_x\xk{\ell}|_{\ell\rightarrow \ell''_c} \rightarrow 0$&  CDM & Yes
\\
RSP$+$CI &$\De_0^0<\De_0^c:\De_0\xk{\ell}|_{\ell\rightarrow +\infty} \rightarrow 0$  & $t^{\prime}|_{\ell\rightarrow +\infty} \rightarrow 0$ & $v_x\xk{\ell}|_{\ell\rightarrow +\infty} \rightarrow +\infty$ & DSM & No
\\\hline
$x$-RVP+CI &$\De_1\xk{\ell}|_{\ell\rightarrow +\infty}=\alpha\xk{\ell}|_{\ell\rightarrow +\infty}=C_1$   &  & &  &
\\
$y$-RVP+CI &$\De_2\xk{\ell}|_{\ell\rightarrow +\infty}=\alpha\xk{\ell}|_{\ell\rightarrow +\infty}=C_2$  & $t^{\prime}|_{\ell\rightarrow +\infty} \rightarrow 0$  & $v_x\xk{\ell}|_{\ell\rightarrow +\infty} \rightarrow v_x^{\ast}$   &RSS& No
\\
RM+CI &$\De_3\xk{\ell}|_{\ell\rightarrow +\infty}=\alpha\xk{\ell}|_{\ell\rightarrow +\infty}=C_3$ & &  &  &
\\ \hline\hline
\end{tabular}
\end{center}
\end{table*}

\section{Summary and discussion \label{Sec:Summ}}

In summary, we have systematically studied the interplay between tilt, disorder, and Coulomb interaction in type-I Dirac fermions. We have shown that the interplay between ordinary tilt and disorder will dynamically generate an anomalous tilt and change the vertex of the Yukawa coupling for Coulomb interaction. After taking these two factors into consideration, we performed extensive RG calculations and find that the Lifshitz transition between two types of Dirac fermions never happens. Without Coulomb interaction, we have found that an arbitrary weak $x$-RVP can induce the diffusive phase transition \cite{Fradkin1, Fradkin2,Shindou09,Ominato14,Goswami11,Kobayashi14,Sbierski14,Roy14,
Syzranov15L,Syzranov15B,Pixley15,Trescher17, Sikkenk17},
which result is never predicted in previous studies for 2D Dirac fermions system.
In addition, we have shown that the interplay between tilt and disorder generates
a bulk nodal arc in the CDM phase. When Coulomb interaction coexists with the disorder,
it generates a critical RSP strength to produce the diffusive phase transition and
ruins the diffusive phase transition caused by $x$-RVP. Instead, a randomly stable
state which is distinct from the DSM and CDM phases appears, and this state is also
present for interplay of Coulomb interaction with $y$-RVP and RM.
All of these results are summarized in Table \ref{Summary}.

Following the custom of previous studies for Coulomb interaction effects to type-I Dirac fermions \cite{Isobe12, ZMHuang17, YWLee18}, we have not explicitly included the Coulomb screening in the derivation of the RG equations. To find out the influence of Coulomb screening, we calculated the vacuum polarization, which takes the form
\bea
\Pi\xk{\om, \b k}=\frac{Ng^2}{16v_x'v_y'}\frac{v_x^{'2}k_x^2+v_y^{'2}k_y^2}
{\sqrt{\xk{\om+it'v_x'k_x}^2+v_x^{'2}k_x^2+v_y^{'2}k_y^2}},\,\,\quad \label{EqPi}
\eea
where $N=2$ is the number of fermion species corresponding to the single Dirac point with two spin components. Comparing with the vacuum polarization provided in Ref. \cite{YWLee18}, the only difference is that the bare tilt and velocities are replaced by the effective ones defined in \Eq{Eqeffquan}. The full propagator of the $\phi$ field $D\xk{\om,\b k}$ is obtained according to the Dyson equation: $D^{-1}\xk{\om,\b k}=D_0^{-1}\xk{0,\b k}+\Pi\xk{\om,\b k}$, which is given by
\bea
 D\xk{\om,\b k}^{-1}&=&\frac{2\alpha v_x\xk{v_x^{'2}k_x^2+v_y^{'2}k_y^2}}
{v_x'v_y'\sqrt{\xk{\om+it'v_x'k_x}^2+v_x^{'2}k_x^2+v_y^{'2}k_y^2}}
 \nn\\&& +
 2\sqrt{k_x^2+k_y^2}  \label{EqFPC}
\eea
Therefore, the screened Coulomb interaction is still long range. This long-range nature is guaranteed by the vanishing $\rho\xk{0}$ for type-I Dirac fermions and it indicates that the effect of the Coulomb interaction cannot change qualitatively by the screening. In addition, remember that $\alpha$ flows to the weak coupling fixed point as $\alpha^{\ast}=\De_i^{\ast} \ll 1$, as a result, $D^{-1}\xk{\om,\b q} \approx D_0^{-1}\xk{0,\b q}$. Therefore, we believe that the analytic results obtained by ignoring the Coulomb screening in Sec.\ref{Sec:IwCi} are still well established when the screening is included.

\emph{Note added.} Recently, we heard of the new work by Yang \cite{Yang2018}, which overlaps somewhat with our work and considers the
correlations of different kinds of disorder from a different perspective.
The main results obtained in our work are quite different
from Ref. \cite{Yang2018}. We have discussed the differences between our work
and Ref. \cite{Yang2018} in Appendix \ref{Sec:Apendiff}.

\section{Acknowledgement}

We would like to thank Z.-K. Yang, J.-R. Wang, and G.-Z. Liu for helpful discussions and comments on the manuscript. The authors acknowledge the financial support provided by Key Research and Development Plan of Ministry of Science and Technology of China under Grant No. 2018YFB1601402.

\appendix

\section{Calculation of RG equations \label{Sec:ApenRG}}

\begin{widetext}

The pertinent one-loop Feynman diagrams are shown in \Fig{FigAll}.
\begin{figure}[htbp]
\center
\subfigure{\includegraphics[width=5.0in]{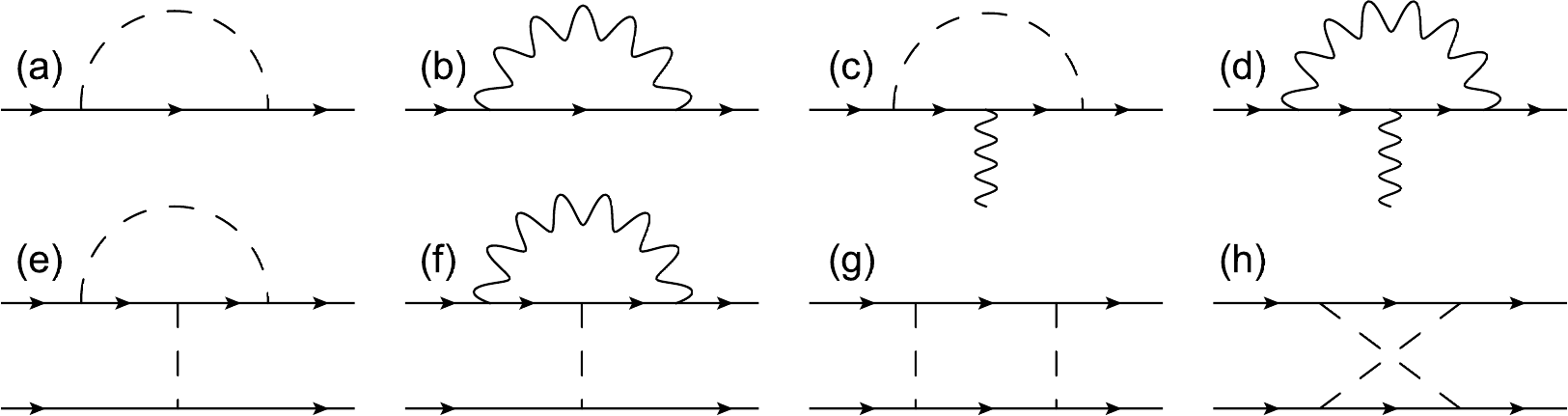}}
\caption{All the related one-loop Feynman diagrams. The solid, dashed, and wavy lines stand for the fermion, disorder, and the Coulomb interaction, respectively.
} \label{FigAll}
\end{figure}

The self-energy correction due to fermion-disorder coupling corresponds to Fig.~\ref{FigAll}(a), and it takes the form
\bea
\Si_{\rm{dis}}^{(a)}\xk{i\om} &=& -\sum_i \kappa_{i}
\int\frac{d^{2}\mathbf{p}}{(2\pi)^{2}}\s_i G_{0}\xk{i\om,\b p}\s_i
\nn\\ &=&
\frac{\xk{\kappa_0+\kappa_1+\kappa_2+\kappa_3}\ell }{2\pi v_x v_y \sqrt{1-t^2}  }\xk{\frac{1-\la t}{1-t^2}}\xk{-i\om \s_0}
+\frac{\xk{\kappa_0+\kappa_1-\kappa_2-\kappa_3}\ell  }{2\pi v_x v_y \sqrt{1-t^2}  }\xk{\frac{1-\la t}{1-t^2}}\xk{i\om t\s_1}.\label{Eqdisseen}
\eea
where the momentum integration is performed by
\bea
\int d^{2}\b p
=\int_{\La e^{-\ell}}^{\La}\abs{\b p}d \abs{\b p}\int_{0}^{2\pi}d\th.
\eea

Next, consider the self-energy coming from Coulomb interaction, which is shown in Fig.~\ref{FigAll}(b), which can be written as
\bea
\Si_C^{(b)}\xk{\b k}
&=& g^{2}\int\frac{d^{2} \b p}{(2\pi)^{2}}\int_{-\infty}^{\infty } \frac{d\omega}{2\pi}\Ga_C
G_{0}\xk{i\omega ,\b{p}}\Ga_CD_0\xk{0,\b k-\b{p}}
\nn \\ &=&
\frac{g^2 f_x \ell }{16\pi v_x} \xk{\la v_x  k_x \s_0}+\frac{g^2 f_x \ell}{16\pi v_x  } \xk{v_x k_x \s_1}+\frac{g^2f_y \ell }{16\pi v_x }\xk{v_y k_y \s_2}
\eea
where $\Ga_C=\s_0+\la\s_1$ represents the effective Coulomb vertex.

Then, considering Fig.~\ref{FigAll}(c), it represents the correction to Yukawa coupling due to disorder and can be written as
\bea
\de g^{(c)}&=&\sum_i \kappa_i  \int\frac{d^{2}\b p}{(2\pi)^{2}}
\s_i G_{0}(0 ,\b{p})\Ga_C G_{0}(0 ,\b{p})\s_i
\nn\\&=&
\frac{\xk{\kappa_0+\kappa_1+\kappa_2+\kappa_3}\xk{1-\la t}\ell \s_0 }{2\pi v_x v_y \xk{1-t^2}^{3/2}  }
-\frac{\xk{\kappa_0+\kappa_1-\kappa_2-\kappa_3}\xk{1-\la t} t \ell\s_1  }{2\pi v_x v_y \xk{1-t^2}^{3/2}  }, \label{Eqgg}
\eea
Next, we consider the correction of Yukawa coupling by Coulomb interaction, which is shown in Fig.~\ref{FigAll}(d), this term just vanishes due to the U(1) gauge invariance. Exactly,  $$\de g^{(d)}=0.$$

Figure \ref{FigAll}(e) represents the correction for disorder vertices caused by the interplay of disorders, it gives rise to
\bea
\delta \kappa_{i}^{(e)} &=& \kappa_{i}\xk{\psi_m^\dagger\s_i\psi_m} \psi_n^\dagger\sum_j \kappa_j \int\frac{d^2\mathbf{p}}{(2\pi)^2} \s_j G_0(0,\mathbf{p})
\s_i  G_0(0,\mathbf{p})\s_j \psi_n,
\eea
For $i=0,1,2,3$, finish the above integration one by one and the results are listed as
\bea
\delta\kappa_i^{(e)}=
\left\{
  \begin{array}{ll}
    \frac{\kappa_0\xk{\sum_{j=0}^3\kappa_j}\ell \xk{\psi_m^\dagger\s_0\psi_m}
    \xk{\psi_n^\dagger\s_0\psi_n}+C_0\xk{\psi^\dagger_{m}\s_1 \psi_{m} }
    \xk{\psi^\dagger_{n}\s_0\psi_{n}}}{2\pi v_x v_y \xk{1-t^2}^{3/2}  }, & \hbox{$i=0$,}
    \\*[0.3cm]
    \frac{t^2\kappa_1\xk{\kappa_0+\kappa_1-\kappa_2-\kappa_3}\ell
    \xk{\psi_m^\dagger\s_1\psi_m}\xk{\psi_n^\dagger\s_1\psi_n}
    +C_1\xk{\psi^\dagger_{m}\s_1 \psi_{m} }
    \xk{\psi^\dagger_{n}\s_0\psi_{n}}}{2\pi v_x v_y \xk{1-t^2}^{3/2}  }, & \hbox{$i=1$,}
    \\
    0, & \hbox{$i=2$,}
    \\
   \frac{-\kappa_3 \xk{\kappa_0-\kappa_1-\kappa_2+\kappa_3}\ell\xk{\psi_m^\dagger\s_3\psi_m}
   \xk{\psi_n^\dagger\s_3\psi_n}  }{2\pi v_x v_y \sqrt{1-t^2}  }, & \hbox{$i=3$.}\label{EqloopC}
  \end{array}
\right.
\eea
We remind here there exist two terms which are proportional to
$\xk{\psi^\dagger_{m}\s_1 \psi_{m} } \xk{\psi^\dagger_{n}\s_0\psi_{n}}$.
However, the correlation between different types of disorder defined in
\Eq{Eqdefdis} forbids the appearance of this cross term in the replica formalism.
Hereafter, we just discard this kind of term, which only appeared in loop corrections.

Then, we take into account Fig.~\ref{FigAll}(f): it stands for the renormalization of disorder couplings due to Coulomb interaction, which is given by
\bea
 \de \kappa_i^{(f)}&=&\xk{\psi^\dagger_{m}\s_i \psi_{m} } \psi^\dagger_{n}\Bigg[  g^2 \int\frac{d^{2}\b p}{(2\pi)^{2}}\int_{-\infty}^{\infty}  \frac{d\omega}{2\pi}
\Ga_C G_{0}(i\omega ,\b{p})\s_i G_{0}(i\omega ,\b{p})\Ga_C D_0(0,\b{p})\Bigg] \psi_{n}
\nn\\&=&
\left\{
\begin{array}{ll}
\frac{\la^2 g^2 f_x\ell \xk{\psi^\dagger_{m}\s_0 \psi_{m} } \xk{\psi^\dagger_{n}\s_0\psi_{n}} }{16\pi v_x \xk{1-\la t}},
& \hbox{$i=0$,}
\\*[0.3cm]
\frac{- g^2 f_x \ell\xk{\psi^\dagger_{m}\s_1 \psi_{m} } \xk{\psi^\dagger_{n}\s_1\psi_{n}}}{16\pi v_x \xk{1-\la t}},
& \hbox{$i=1$,}
\\*[0.3cm]
\frac{- g^2f_y \ell \xk{\psi^\dagger_{m}\s_2 \psi_{m} } \xk{\psi^\dagger_{n}\s_2\psi_{n}}}{16\pi v_x}, & \hbox{$i=2$,}
\\*[0.3cm]
\frac{- g^2 \zk{ \xk{1-\la^2}f_x+\xk{1-\la t} f_y }\ell \xk{\psi^\dagger_{m}\s_3 \psi_{m} } \xk{\psi^\dagger_{n}\s_3\psi_{n}}}{16\pi v_x\xk{1-\la t} },  & \hbox{$i=3$.}
\end{array}
\right.
\eea

At last, we compute the disorder vertices corrections coming from the sum of ZS'- and BCS-type graphs, exactly Figs.~\ref{FigAll}(g)+\ref{FigAll}(h):
\bea
\delta\kappa^{(g)+(h)} &=& \sum_{ij}\kappa_i \kappa_j\int\frac{d^{2}\mathbf{p}}{(2\pi)^{2}}\psi^{\dagger}_m\l[\s_i G_0(0,\mathbf{p})\s_j\rt]\psi_m
\psi^{\dagger}_n\l[\s_j G_0(0,\mathbf{p})\s_i+\s_i G_0(0,-\mathbf{p})\s_j\rt]\psi_n,\label{Eqe+f}
\eea
As a result, $\delta\kappa^{(g)+(h)}=0$ for $\s_i=\s_j$, due to $G_0(0,-\mathbf{p})=-G_0(0,\mathbf{p})$. Therefore, we only need to consider $\s_i\neq\s_j$, which contains six pairs as
$\l(\s_i,\s_j\rt)= \{\xk{\s_0,\s_1}, \xk{\s_0,\s_2}, \xk{\s_0,\s_3}, \xk{\s_1,\s_2}, \xk{\s_1,\s_3}, \xk{\s_2,\s_3}\}$. After calculating them one by one, the results can be written as
\bea
\delta
\kappa^{(g)+(h)}=
\left\{
\begin{array}{ll}
\frac{-\kappa_0 \kappa_1\ell\xk{\psi^{\dagger}_m\s_3\psi_m} \xk{\psi^{\dagger}_n\s_3\psi_n}}{2\pi v_xv_y\sqrt{1-t^2}},
& \hbox{$\l(\s_i,\s_j\rt)= \xk{\s_0,\s_1}$,}
\\*[0.3cm]
\frac{-\kappa_0 \kappa_2 \ell\xk{\psi^{\dagger}_m\s_3\psi_m}
\xk{\psi^{\dagger}_n\s_3\psi_n}}{2\pi v_xv_y\xk{1-t^2}^{3/2}},
& \hbox{$\l(\s_i,\s_j\rt)= \xk{\s_0,\s_2}$,}
\\*[0.3cm]
\frac{-\kappa_0 \kappa_3 \ell\zk{\xk{\psi^{\dagger}_m\s_2\psi_m} \xk{\psi^{\dagger}_n\s_2\psi_n}-\xk{1-t^2}\xk{\psi^{\dagger}_m\s_1\psi_m} \xk{\psi^{\dagger}_n\s_1\psi_n} }  }{2\pi v_xv_y\xk{1-t^2}^{3/2}},
& \hbox{$\l(\s_i,\s_j\rt)= \xk{\s_0,\s_3}$,}
\\*[0.3cm]
\frac{-t^2\kappa_1\kappa_2 \ell\xk{\psi^{\dagger}_m\s_3\psi_m}\xk{\psi^{\dagger}_n\s_3\psi_n}}{2\pi v_xv_y\xk{1-t^2}^{3/2}},
& \hbox{$\l(\s_i,\s_j\rt)= \xk{\s_1,\s_2}$,}
\\*[0.3cm]
\frac{-\kappa_1 \kappa_3 \ell\zk{t^2\xk{\psi^{\dagger}_m\s_2\psi_m} \xk{\psi^{\dagger}_n\s_2\psi_n}-\xk{1-t^2}\xk{\psi^{\dagger}_m\s_0\psi_m} \xk{\psi^{\dagger}_n\s_0\psi_n}  }  }{2\pi v_xv_y\xk{1-t^2}^{3/2}},
& \hbox{$\l(\s_i,\s_j\rt)= \xk{\s_1,\s_3}$,}
\\*[0.3cm]
\frac{-\kappa_2 \kappa_3 \ell \zk{t^2\xk{\psi^{\dagger}_m\s_1\psi_m} \xk{\psi^{\dagger}_n\s_1\psi_n}-\xk{\psi^{\dagger}_m\s_0\psi_m} \xk{\psi^{\dagger}_n\s_0\psi_n}  }}{2\pi v_xv_y\xk{1-t^2}^{3/2}},
& \hbox{$\l(\s_i,\s_j\rt)= \xk{\s_2,\s_3}$.}
\end{array}
\right.
\eea

After finishing the calculations of loop corrections, one can insert all the corrections into the free action, and make the following rescaling:
\bea
\tilde{\om}&=&b\om,\quad \tilde{k}_i=bk_i,\quad\tilde{\psi}=\sqrt{Z_{\psi}}\psi,\quad
\tilde{t}=Z_{t} t,\quad\tilde{\la}=Z_{\la} \la,\quad
\tilde{\phi}=e^{-2\ell}\phi,
\nn \\
\tilde{v}_{x}&=&Z_{v_x}v_x,\quad
\tilde{v}_{y}=Z_{v_y}v_y,\quad\tilde{\kappa}_i=Z_{\kappa_i}\kappa_i,
\quad\tilde{g}=Z_{g}g,\quad (b\equiv e^{\ell}),
\eea
then the action becomes
\bea
\tilde{S} &=&
\int \frac{d^2\mathbf{\tilde{k}}d\tilde{\omega}}{(2\pi)^3}
\frac{e^{-3\ell}\tilde{\psi}^{\dag}_m}{Z_{\psi}}\Bigg\{-i\tilde{\om }e^{-\ell}\Bigg[\s_0\xk{1+\frac{\xk{\sum_i\kappa_i}\xk{1-\la t  } \ell  }{2\pi v_x v_y \xk{1-t^2}^{3/2}  }   }
+\frac{\tilde{\la}}{Z_{\la}} \s_1 \Bigg(1- \frac{t\xk{1-\la t} }{\la}
\frac{\xk{\kappa_0+\kappa_1-\kappa_2-\kappa_3}\ell  }{2\pi v_x v_y \xk{1-t^2}^{3/2}  }    \Bigg)     \Bigg]
\nn\\&&
+\frac{\tilde{v}_x \tilde{t}e^{-\ell} \tilde{k}_x\s_0}{Z_{v_x}Z_{t}} \xk{1-\frac{g^2f_x \la \ell }{16 \pi v_x t } }
+\frac{\tilde{v}_xe^{-\ell} \tilde{k}_x \s_1}{Z_{v_x}}\xk{1+\frac{g^2 f_x \ell  }{16 \pi v_x }  }+
\frac{\tilde{v}_y\tilde{k}_y e^{-\ell} \s_2}{Z_{v_y}} \xk{1+\frac{g^2 f_y \ell }{16 \pi v_x} } \Bigg\} \tilde{\psi}_m
\nn\\&&
-\int\frac{d\tilde{\omega}_1
d\tilde{\omega}_2d^{2}\mathbf{\tilde{k}}_1d^{2}\mathbf{\tilde{k}}_2d^{2}\mathbf{\tilde{k}}_3}
{(2\pi)^{8}}\frac{e^{-8\ell}}{Z_{\psi}^2}\Bigg\{\frac{\tilde{\kappa}_0}
{2Z_{\kappa_0}}\tilde{\psi}^\dagger_{m}(i\tilde{\omega}_1,\mathbf{\tilde{k}}_1)
\s_0
\tilde{\psi}_{m}(i\tilde{\omega}_1,\mathbf{\tilde{k}}_2)
\tilde{\psi}^\dagger_{n}(i\tilde{\omega}_2,\mathbf{k}_3)\s_0\tilde{\psi}_{n}
\xk{i\tilde{\omega}_2,\sum_{i=1}^{3}\mathbf{\tilde{k}}_i}
\nn \\ &&\times
\Bigg[1+\frac{2\xk{\sum_i\kappa_i}\ell  }{2\pi v_x v_y \sqrt{1-t^2}  }-\frac{2 g^2 f_x\la^2 \ell }{16\pi v_x\xk{1-\la t}}
+\frac{2\xk{\kappa_1+\frac{\kappa_2}{1-t^2}} \kappa_3 \ell}{2\pi v_xv_y\sqrt{1-t^2} \kappa_0 } \Bigg]
+\frac{\tilde{\kappa}_1}{2Z_{\kappa_1}}\tilde{\psi}^\dagger_{m}
(i\tilde{\omega}_1,\mathbf{\tilde{k}}_1)\s_1
\tilde{\psi}_{m}(i\tilde{\omega}_1,\mathbf{\tilde{k}}_2)
\nn\\&&\times
\tilde{\psi}^\dagger_{n}(i\tilde{\omega}_2,\mathbf{k}_3)\s_1
\tilde{\psi}_{n}\xk{i\tilde{\omega}_2,\sum_{i=1}^{3}\mathbf{\tilde{k}}_i}
\zk{1+\frac{2t^2\xk{\kappa_0+\kappa_1-\kappa_2-\kappa_3}\ell  }{2\pi v_x v_y \xk{1-t^2}^{3/2}  }+ \frac{2 g^2 f_x \ell }{16\pi v_x\xk{1-\la t}}+ \frac{2 \kappa_3 \xk{\frac{t^2\kappa_2}{1-t^2}+\kappa_0} \ell}{2\pi v_xv_y\sqrt{1-t^2} \kappa_1}  }
\nn\\&&
+\frac{\tilde{\kappa}_2}{2Z_{\kappa_2}}\tilde{\psi}^\dagger_{m}
(i\tilde{\omega}_1,\mathbf{\tilde{k}}_1)\s_2
\tilde{\psi}_{m}(i\tilde{\omega}_1,\mathbf{\tilde{k}}_2)
\tilde{\psi}^\dagger_{n}(i\tilde{\omega}_2,\mathbf{k}_3)\s_2\tilde{\psi}_{n}
\xk{i\tilde{\omega}_2,\sum_{i=1}^{3}\mathbf{\tilde{k}}_i}
\zk{1+ \frac{2g^2f_y \ell}{16\pi v_x } +\frac{2 \kappa_3\xk{t^2\kappa_1+\kappa_0} \ell}{2\pi v_xv_y\xk{1-t^2}^{3/2} \kappa_2} }
\nn\\&&
+\frac{\tilde{\kappa}_3}{2Z_{\kappa_3}}\tilde{\psi}^\dagger_{m}
(i\tilde{\omega}_1,\mathbf{\tilde{k}}_1)\s_3
\tilde{\psi}_{m}(i\tilde{\omega}_1,\mathbf{\tilde{k}}_2)
\tilde{\psi}^\dagger_{n}(i\tilde{\omega}_2,\mathbf{k}_3)\s_3\tilde{\psi}_{n}
\xk{i\tilde{\omega}_2,\sum_{i=1}^{3}\mathbf{\tilde{k}}_i}
\Bigg[1-\frac{2\xk{\kappa_0-\kappa_1-\kappa_2+\kappa_3}\ell  }{2\pi v_x v_y \sqrt{1-t^2}}
\nn\\&&+
\frac{2g^2\ell}{16\pi v_x}\xk{ \frac{1-\la^2}{1-\la t} f_x+f_y }
+ \frac{2 \zk{\kappa_0\kappa_1+\frac{ \xk{t^2\kappa_1+\kappa_0}\kappa_2  }{1-t^2}}\ell}{2\pi v_xv_y\sqrt{1-t^2} \kappa_3}  \Bigg] \Bigg\}+
\int \frac{d^2\mathbf{\tilde{k}}d\tilde{\omega}}{(2\pi)^3} \frac{d^2\mathbf{\tilde{k}'}d\tilde{\omega}'}{(2\pi)^3}
\frac{i\tilde{g}}{Z_{g}Z_{\psi}}
\tilde{\psi}^\dagger_m\xk{\tilde{\om},\b{\tilde{k}} }
\nn \\ &&\times
 e^{2\ell}\tilde{\phi}\xk{\tilde{\om}-\tilde{\om}',\b{\tilde{k}}-\b{\tilde{k}'} } \Bigg\{
\frac{\tilde{\la}\s_1}{Z_{\la}}
\zk{1- \frac{t\xk{1-\la t}}{\la}
\frac{\xk{\kappa_0+\kappa_1-\kappa_2-\kappa_3}\ell  }{2\pi v_x v_y \xk{1-t^2}^{3/2}  }  }
+\s_0\zk{1+\frac{\xk{\sum_i\kappa_i}\ell  }{2\pi v_x v_y \sqrt{1-t^2}  }\xk{\frac{1-\la t}{1-t^2}}   }  \Bigg\}
\nn\\&&\times \tilde{\psi}_m\xk{\tilde{\om}',\b{\tilde{k}'}  } +
\frac{1}{2}\int \frac{d^2\mathbf{\tilde{k}}d\tilde{\omega}}{(2\pi)^3}e^{-3\ell}e^{2\ell}\phi\xk{\tilde{\om},\b{ \tilde{k}} }e^{\ell} D_{0}^{-1}\xk{\tilde{\om},\b{\tilde{k}}}  \tilde{\phi}\xk{\tilde{\om},\b{\tilde{k}}}.
\eea
By requiring this action takes the same form of the original one, we obtain the following renormalization factors
\bea
Z_{\la}&=&
1-\frac{\xk{\kappa_0+\kappa_1-\kappa_2-\kappa_3}\xk{1-\la t}\ell  }{2\pi v_x v_y \xk{1-t^2}^{3/2}  }\xk{\frac{t }{\la } }-\frac{\xk{\sum_i\kappa_i}\xk{1-\la t}\ell  }{2\pi v_x v_y \xk{1-t^2}^{3/2}  },
\\
Z_{v_x}&=&1+\frac{g^2 f_x }{16 \pi v_x}\ell-\frac{\xk{\sum_i\kappa_i}\xk{1-\la t}\ell  }{2\pi v_x v_y \xk{1-t^2}^{3/2}  },
\\
Z_{t}&=&1-\frac{g^2  f_x }{16 \pi v_x}\xk{ 1-\frac{\la}{t}}\ell,
\\
Z_{v_y}&=&1+\frac{g^2 f_y }{16 \pi v_x}\ell -\frac{\xk{\sum_i\kappa_i}\xk{1-\la t}\ell  }{2\pi v_x v_y \xk{1-t^2}^{3/2}  } ,
\\
Z_g &=&1,
\\
Z_{\De_0}&=&1+\frac{2\xk{\sum_i\kappa_i} \ell  }{2\pi v_x v_y \xk{1-t^2}^{3/2} } +
\frac{2\xk{\kappa_1+\frac{\kappa_2}{1-t^2}} \kappa_3 \ell}{2\pi v_xv_y\sqrt{1-t^2} \kappa_0 }
-\frac{2 g^2 f_x\la^2\ell }{16\pi v_x\xk{1-\la t}}
-\frac{2\xk{\sum_i\kappa_i}\xk{1-\la t}  \ell }{2\pi v_x v_y \xk{1-t^2}^{3/2}  },
\\
Z_{\kappa_1}&=&
1+\frac{2t^2\xk{\kappa_0+\kappa_1-\kappa_2-\kappa_3}\ell  }{2\pi v_x v_y \xk{1-t^2}^{3/2}  }
+ \frac{2 g^2 f_x\ell }{16\pi v_x\xk{1-\la t}}+ \frac{2 \kappa_3\xk{\frac{t^2\kappa_2}{1-t^2}+\kappa_0} \ell}{2\pi v_xv_y \sqrt{1-t^2} \kappa_1}
-\frac{2\xk{\sum_i\kappa_i}\xk{1-\la t} \ell  }{2\pi v_x v_y \xk{1-t^2}^{3/2}  } ,
\\
Z_{\kappa_2}&=&
1+ \frac{2g^2 f_y\ell }{16\pi v_x } +\frac{2 \kappa_3 \xk{t^2\kappa_1+\kappa_0}\ell}{2\pi v_xv_y\xk{1-t^2}^{3/2} \De_2}-\frac{2\xk{\sum_i\kappa_i}\xk{1-\la t} \ell  }{2\pi v_x v_y \xk{1-t^2}^{3/2}  } ,
\eea
\bea
Z_{\De_3}&=&1-\frac{2\xk{\kappa_0-\kappa_1-\kappa_2+\kappa_3}\ell  }{2\pi v_x v_y \sqrt{1-t^2}}-\frac{2\xk{\sum_i\kappa_i} \ell  }{2\pi v_x v_y }\frac{\xk{1-\la t}}{\xk{1-t^2}^{3/2}}
+\frac{2g^2\ell}{16\pi v_x }\xk{ \frac{1-\la^2}{1-\la t} f_x+f_y }
\nn\\&&+
\frac{2\zk{\kappa_0\xk{\kappa_1+\kappa_2}
+t^2\kappa_1\xk{\kappa_2-\kappa_0} }\ell }{2\pi v_xv_y\xk{1-t^2}^{3/2}\kappa_3}.
\eea

To proceed, we simplify above expressions by making the following redefinitions of the couplings as
\bea
\frac{\kappa_i }{2\pi v_x v_y \sqrt{1-t^2}  } \equiv \De_i,\quad \frac{g^2}{16\pi v_x} \equiv \alpha.
\eea
Where the rescaled $\De_i$ represent the effective disorder strength and $\alpha$ is the effective Coulomb interaction. The beta function can be obtained by the renormalization factors according to
\bea
\frac{d\ln X}{d\ell}=\frac{d Z_X}{d\ell}\bigg|_{\ell=0}.
\eea
Then, the above renormalization factors produce the RG equations with rescaled couplings as are given in the main text.
\end{widetext}

\section{DOS and Specific heat  \label{Sec:Apenob}}

\subsection{Density of states}

The DOS $\rho(\omega)$ is defined by
\begin{eqnarray}
\rho(\omega)=-N\int \frac{d^2 \mathbf{k}}{(2 \pi)^2} \trace{\Im{G^{R}(\omega,k_x,k_y)}},\label{EqDOSexpr}
\end{eqnarray}
where $N=2$, represents the flavor of Dirac
fermions. After performing a standard calculation \cite{Colemanb} by using the amended action of \Eq{Eqaction}, we found
\bea
\rho_0(\omega)&=&\frac{\abs{\om}\xk{1-t\la}^2}{\pi\xk{1-\la^2}  v_y v_x \left(1-t^{2}\right){}^{3/2}}\label{EqDOSfree}
\eea
The interactions can be divided into two classes according to whether it induces the DSM-CDM phase transition or not. Following the line of our main text, we firstly focus on the effect of RSP or $x$-RVP without Coulomb interaction. According to the calculation performed in the main text, at zero-energy, the self-energy induced by RSP or $x$-RVP in the formalism of self-consistent Born approximation is given by
\bea
\Si^{\text{SCBA} }\xk{\om=0}=-i\ga_{0j}\xk{\s_0-t\s_1},
\eea
where $j=0,1$ and in general, $\ga_{00}\neq \ga_{01}$. By including the above self-energy, the retarded Green function reads as,
\bea
 G^{R}(\omega=0,k_x,k_y)&=&-[v_x t_0 k_x\s_0 +v_x k_x \s_1+v_y k_y \s_2\nn\\&&-i\ga_{0j}\xk{\s_0-t\s_1}  ]^{-1}.
\eea
After substituting this result into \Eq{EqDOSexpr}, one will find that the momentum integral is divergent \cite{Fradkin2}.  To control it, we introduce a UV cutoff $\La$ which is determined by the bandwidth, and perform the momentum integration by $\int d^2\b k=\int_{\ga_0}^{\La}\abs{\b k}d\abs{\b k} \int_{0}^{2\pi}d\th$, then the zero-energy DOS is found to be
\bea
\rho\xk{0}\sim \frac{\xk{1+t_0^2}\ga_{0j}\ln\xk{\La/\ga_{0j}}}{\xk{1-t_0^2}^{3/2}}.
\eea
This nonzero $\rho(0)$ exists only in CDM phase.

When considering the influence of interactions that do not induce the phase transition, their effects are embodied in the corrections to velocities and tilts.  To proceed, we employ the transformation $\om = \La e^{-\ell}$, where $\om$ is the low-energy scale we are concerned with. According to \Eq{EqDOSfree}, we have
\bea
\frac{d\ln \rho\xk{\om}}{d\ln \om}&=&1+\frac{2d\ln \xk{1-t\la}}{d\ln \om}-\frac{d\ln v_x}{d\ln \om}-\frac{d\ln v_y}{d\ln \om}
\nn\\&&
-\frac{3}{2}\frac{ d\ln \xk{1-t^2} }{d \ln \om}\label{EqDOSom}.
\eea

(i) When there is only $y$-RVP, $\De_{2}\xk{\om}=\De_{2}^0$, we obtained
\bea
\rho\xk{\om}=\rho\xk{\La}\xk{\frac{\om}{\La}}^{\displaystyle 1-2\De_{2}^0\xk{1+t_0^2}/\xk{1-t_0^2}},
\eea
where $\rho\xk{\La}$ is a constant density of states at the scale of UV cutoff. In low energy where $\om \ll \La$,
\bea
\rho\xk{\omega} \sim \om^{\displaystyle 1-2\De_{2}^0\xk{1+t_0^2}/\xk{1-t_0^2}}.
\eea
Therefore, the marginal $y$-RVP provides a power-law enhancement to the low-energy DOS and the nonzero tilt is helpful to increase the exponent generated by $y$-RVP.

(ii) When only RM is present, there is $$\De_3\xk{\om}=\De_3^0/\zk{1+2\De_3^0\ln\xk{\La/\om} },$$ with $\De_3^0=\De_3\xk{\La}$,
we obtained,
\bea
\rho\xk{\om}&=&\rho\xk{\La}\xk{\frac{\om}{\La}}\zk{1+2\De_3^0\ln\xk{\La/\om}}.
\eea
In low energy,
\bea
\rho\xk{\omega} \sim \om \ln \om,
\eea
which means the marginally irrelevant RM supplies a logarithmic enhancement to the low-energy DOS.

(iii) For clean limit without velocity anisotropy, we have
\bea
\alpha\xk{\om}=\frac{\alpha_0}{1+\alpha_0\ln\xk{\La/\om} },\quad t\xk{\om}=\frac{t_0}{1+\alpha_0\ln\xk{\La/\om} },
\nn\\
\eea
and the DOS is given by
\bea
\rho\xk{\om}=\frac{ \rho\xk{\La}\zk{1+\alpha_0\ln\xk{\La/\om} }\xk{1-t_0^2}^{3/2} \xk{\om/\La} }{ \zk{\zk{1+\alpha_0\ln\xk{\La/\om} }^2-t_0^2 }^{3/2}  },\qquad
\eea
In low energy,
\bea
\rho\xk{\omega} \sim \frac{\om }{\ln^2 \om},
\eea
which indicates the marginally irrelevant Coulomb interaction contributes a logarithmic suppression to DOS.

(iv) For the interplay of Coulomb interaction and RVP or RM, the analytical expression for $\rho\xk{\om}$ is impossible to find due to the existence of too many parameters. However, in low energy, we can replace all the parameters by their fixed points, which permits us to obtain the behavior of $\rho\xk{\om}$ in low energy. Consider the coexistence of $x$-RVP and Coulomb interaction, according to \Eq{EqDOSom}, the low-energy DOS takes the form
\bea
\frac{d\ln \rho\xk{\om}}{d\ln \om}\bigg|_{\om\ll\La}&=&1-\frac{2\De_1^{\ast}\xk{1+ t_{\ast}^2} }{1-t_{\ast}^2}+\alpha_{\ast} f_y^{\ast}
+\alpha_{\ast} f_x^{\ast}
\nn\\&&\times\zk{\frac{2\xk{t_{\ast}-\la_{\ast}}^2}{\xk{1-t_{\ast}^2}\xk{1-t_{\ast}\la_{\ast}}}
+\frac{1-t_{\ast}\la_{\ast}}{1-t_{\ast}^2} }
\nn\\&=&
1.
\eea
Similarly, when Coulomb interaction coexists with $y$-RVP or RM, the low-energy DOS can be written as,
\bea
\frac{d\ln \rho\xk{\om}}{d\ln \om}\bigg|_{\om\ll\La}&=&
1-2\De_i^{\ast}+\alpha_{\ast} f_y^{\ast}+\alpha_{\ast} f_x^{\ast}
\nn\\&& \times
\zk{\frac{2\xk{t_{\ast}-\la_{\ast}}^2}{\xk{1-t_{\ast}^2}\xk{1-t_{\ast}\la_{\ast}}} +\frac{1-t_{\ast}\la_{\ast}}{1-t_{\ast}^2} }
\nn\\&=&1,
\eea
where $i=2, 3$ represents $y$-RVP and RM, respectively. Therefore, for the coexistence of Coulomb interaction and RVP or RM, the low-energy DOS behaves as
\bea
\rho\xk{\om}\sim \om,
\eea
which has the same exponent as the free DOS.

\subsection{Specific heat}

To calculate the specific heat, we need to compute the free energy firstly.  After performing the functional integration by the standard way as shown closely in \cite{Colemanb}, the free energy of fermions is given by
\begin{eqnarray}
F_{f}(T)=-2NT\sum_{\omega_{n}}\!\! \int\frac{d^2\mathbf{p}}{(2\pi)^2}\!\!
\ln\abs{\Det{G^{-1}\xk{\om_n,\b p}}},\qquad
\end{eqnarray}
where $\omega_{n}=(2n+1)\pi T$ is the Matsubara frequency. Finishing the frequency summation and momentum integral yields
\bea
F_{f}(T) =
-\frac{6 \zeta (3)\xk{1-t\la}^2}{2\pi v_y v_x \xk{1-t^2}^{3/2}}T^3,
\eea
where $\zeta (x)$ is the Riemann zeta function. The specific heat can be obtained by
\begin{eqnarray}
C_{v}(T) = -T\frac{\partial^2 F_{f}(T)}{\partial T^2} =
\frac{18 \zeta (3)\xk{1-t\la}^2T^2}{\pi v_y v_x \xk{1-t^2}^{3/2}}.
\end{eqnarray}

The corrections to specific heat from interactions are calculated similarly as DOS, we just summarize the results here
\begin{widetext}
\bea
C_{v}(T) \sim
\left\{
\begin{array}{ll}
\frac{\left(t_0^2+1\right)\ga_{0j}\ln\xk{\La/\ga_{0j}} T}{\left(1-t_0^2\right)^{3/2} },
& \hbox{Only $x$-RVP or RSP,}
\\*[0.3cm]
C_v\xk{T_{\La}}\xk{\frac{T}{T_{\La}}}^{2-2\De_{2}^0\xk{1+t_0^2}/\xk{1-t_0^2}},
& \hbox{Only $y$-RVP,}
\\*[0.3cm]
C_v\xk{T_{\La}}\xk{\frac{T}{T_{\La}}}^2\zk{1+2\De_3^0\ln\xk{T_{\La}/T}},
& \hbox{Only RM,}
\\*[0.3cm]
\frac{ C_v\xk{T_{\La}}\zk{1+\alpha_0\ln\xk{\frac{T_{\La}}{T}} }\xk{1-t_0^2}^{3/2} }{ \zk{\zk{1+\alpha_0\ln\xk{\frac{T_{\La}}{T}} }^2-t_0^2 }^{3/2}  },
& \hbox{Only Coulomb}
\\*[0.3cm]
T^2,
& \hbox{Coulomb $+$ $y$-RVP or RM.}
\end{array}
\right.
\eea
\end{widetext}
Here, $T_{\La}$ is certain fixed high temperature and $T = T_{\La}e^{-\ell}$.

\section{Differences between our work and Ref. \cite{Yang2018} \label{Sec:Apendiff}}

First, in Ref. \cite{Yang2018}, neither RSP nor $x$-RVP can exist alone
in the system because they both dynamically generates other types of disorder.
In consequence, the system inevitably contains five types of disorder.
Therefore, the influence of RSP or $x$-RVP on tilted Dirac semimetals
cannot be identified in Ref. \cite{Yang2018}. In our paper, any type of
disorder could exist alone in tilted Dirac semimetal, which permits us to
study the effect of every single type of disorder.

Second, in Ref. \cite{Yang2018}, the effective tilt approaches to $1$
when the DM-CDM phase transition happens while it's a small constant in our work.

Third, the low-energy DOS we obtained when $y$-RVP coexists with Coulomb
interaction is different from the one given in Ref. \cite{Yang2018}. So does for RM.

Moreover, we presented some results which have not been considered in
Refs. \cite{Yang2018}. In particular, we have provided a detailed calculation
for the retarded fermion self-energy by means of SCBA to show the formation
of the bulk nodal arc. We have clarified that the origin of the bulk nodal
arc in a tilted Dirac semimetal with particle-hole symmetry is different
from the one found in a tilted Dirac semimetal without any external symmetry.
We have verified that the tilt breaks the time-independent gauge symmetry,
which induces $x$-RVP to become marginally relevant.

At last, we show that there exist two potential mistakes in Refs. \cite{Yang2018}. One mistake involves Eq.(A18) in Ref. \cite{Yang2018}, which takes the form
\bea
\big(\psi^{\dagger}\!\!\!\!\!\!&&\!\!\!\!\!\!\sigma_{1} \psi\big)\left(\psi^{\dagger} \sigma_{0} \psi\right)
\nn\\ \!\!&=&\!\!
-2\left(\psi^{\dagger} \sigma_{-} \psi\right)\left(\psi^{\dagger} \sigma_{-} \psi\right)
+\frac{1}{2}\left(\psi^{\dagger} \sigma_{0} \psi\right)\left(\psi^{\dagger} \sigma_{0} \psi\right)
\nn\\&&\!\!
+\frac{1}{2}\left(\psi^{\dagger} \sigma_{1} \psi\right)\left(\psi^{\dagger} \sigma_{1} \psi\right), \label{Eqgene}
\eea
where $\sigma_{-}=\frac{1}{2}\left(\sigma_{0}-\sigma_{1}\right)$.
Based on \Eq{Eqgene}, the authors in Ref. \cite{Yang2018}
claimed that disorder defined by
\bea
S_{-}^{\mathrm{dis}}=\frac{\Delta_{-}}{2} \int d \tau d \tau^{\prime}
d^{2} \mathbf{x}\left(\psi_{a}^{\dagger} \sigma_{-}
\psi_{a}\right)_{\tau}\left(\psi_{b}^{\dagger} \sigma_{-}
\psi_{b}\right)_{\tau^{\prime}} \label{Eqde-}
\eea
is dynamically generated even if it is not present at the beginning. We remind that \Eq{Eqgene} only establishes when $\left(\psi^{\dagger} \sigma_{1} \psi\right)$ and $\left(\psi^{\dagger} \sigma_{0} \psi\right)$ are located at the same space-time point. Once this condition is not satisfied, we obtain
\bea
-\!\!\!&2&\!\!\!\xk{\psi^{\dagger} \sigma_{-} \psi}_x\xk{\psi^{\dagger} \sigma_{-} \psi }_{x'}
+\frac{1}{2}\xk{\psi^{\dagger} \sigma_{0}}_x\xk{\psi^{\dagger} \sigma_{0} \psi}_{x'}
\nn\\&&\!\!
+\frac{1}{2}\xk{\psi^{\dagger}\sigma_{1} \psi}_{x}\xk{\psi^{\dagger} \sigma_{1} \psi}_{x'}
\nn\\\!\!&=&\!\!
\frac{1}{2}\zk{\xk{\psi^{\dagger} \sigma_{1} \psi}_x\xk{\psi^{\dagger} \sigma_{0} \psi}_{x'}+\xk{\psi^{\dagger} \sigma_{0} \psi}_x\xk{\psi^{\dagger} \sigma_{1} \psi}_{x'}}
\nn\\\!\!&\neq&\!\! \xk{\psi^{\dagger}\sigma_{1} \psi}_x\xk{\psi^{\dagger} \sigma_{0} \psi}_{x'},
\eea
where $x\neq x'$ is assumed. This result tells us that \Eq{Eqgene} cannot apply to disorder coupling. Therefore, taking \Eq{Eqgene} as the basis to claim that the disorder described by \Eq{Eqde-} is dynamically generated is a mistake. In fact, even if one assumes that \Eq{Eqgene} could apply to disorder (although this is not true), there still exists another problem. To see this, we generalize \Eq{Eqgene}, which gives rise to
\bea
\big(\psi^{\dagger}\!\!\!\!\!\!&&\!\!\!\!\!\!\sigma_1 \psi\big)\left(\psi^{\dagger} \sigma_{0} \psi\right)
\nn\\\!\!&=&\!\!
\frac{1}{2ab}\bigg[\left(\psi^{\dagger} \xk{a\s_0+b\s_x} \psi\right)\left(\psi^{\dagger} \xk{a\s_0+b\s_x} \psi\right)
\nn\\&&\!\!
-a^2\left(\psi^{\dagger} \sigma_{0} \psi\right)\left(\psi^{\dagger} \sigma_{0} \psi\right)
-b^2\left(\psi^{\dagger} \sigma_1 \psi\right)\left(\psi^{\dagger} \sigma_1 \psi\right) \bigg], \qquad \label{Eqdecg}
\eea
where $a$ and $b$ are two real constants with nonzero values. If we take $a=1/2$ and $b=-1/2$, then \Eq{Eqdecg} becomes \Eq{Eqgene}. By taking $a=1/2$ and $b=1/2$, we obtain
\bea
\big(\psi^{\dagger}\!\!\!\!\!\!&&\!\!\!\!\!\! \sigma_1 \psi\big)\left(\psi^{\dagger} \sigma_{0} \psi\right)
\nn\\\!\!&=&\!\!
2\left(\psi^{\dagger} \frac{1}{2}\xk{\s_0+\s_x} \psi\right)\!\left(\psi^{\dagger} \frac{1}{2}\xk{\s_0+\s_x} \psi\right)
\nn\\&&\!\!
-\frac{1}{2}\!\left(\psi^{\dagger} \sigma_{0} \psi\right)\!\left(\psi^{\dagger} \sigma_{0} \psi\right)\!-\!\frac{1}{2}\!\left(\psi^{\dagger} \sigma_1 \psi\right)\!\left(\psi^{\dagger} \sigma_1 \psi\right).\label{Eqgene+}
\eea
According to the argument in Ref. \cite{Yang2018}, the disorder described by \Eq{Eqde-} is dynamically generated due to \Eq{Eqgene}.
If this is true, another disorder defined by the following action would also be dynamically generated
\bea
S_{+}^{\mathrm{dis}}=\frac{\Delta_{+}}{2} \int d \tau d \tau^{\prime} d^{2} \mathbf{x}\left(\psi_{a}^{\dagger} \Gamma_{+} \psi_{a}\right)_{\tau}\left(\psi_{b}^{\dagger} \Gamma_{+} \psi_{b}\right)_{\tau^{\prime}}, \label{Eqgenes+}
\eea
where $\Gamma_{+}=\frac{1}{2}\left(\sigma_{0}+\sigma_1\right)$.
According to \Eq{Eqdecg}, an infinite number of types
of disorder are dynamically generated by taking different values of $a$ and $b$.
This appears to be inaccurate because there is no system which
could support an infinite number of interactions.
The appearance of this result stems from the misuse of \Eq{Eqgene}, which enables us to confirm that the consideration of dynamically generated disorder in Ref. \cite{Yang2018} is indeed a mistake.

Another mistake is related to the polarization function, which is given by Eq.(7) in Ref. \cite{Yang2018}. In Eq.(7) of Ref. \cite{Yang2018}, the polarization function
depends on the bare fermion velocities and tilt. However, the effective fermion
velocities and tilt are the really physical observables.
A correct polarization function should be a function of the effective fermion
velocities and tilt, which is given by \Eq{EqPi} in our work.
The authors in Ref. \cite{Yang2018} employed an incorrect polarization function to
analyze the interplay between disorder and Coulomb interaction,
and hence the conclusions in Sec.C of Ref. \cite{Yang2018} appear to be not credible.


\bibliographystyle{apsrev4-1}
\bibliography{Tilt_coul_dis_refs}

\end{document}